\documentstyle[preprint,aps]{revtex}
\input epsfig.sty
\def\qu{\quad}
\def\scr{\scriptstyle}
\def\sscr{\scriptscriptstyle}
\def\imp{{{}\ifmmode\;\Rightarrow\;\else$\;\Rightarrow\;$ \fi}}

\def\beg{\begin{equation}}
\def\eb{\end{equation}}
\def\eqn#1{(\ref{#1})}
\def\dis{\displaystyle}
\def\txt{\textstyle}
\def\intl{\displaystyle\int\limits}
\def\Sum{\dis\sum}

\font\rbig=cmr12 scaled\magstep1
\font\med=cmr12 scaled\magstep0

\def\<{\noindent }
\def\rto{\longrightarrow}
\def\lrto#1{\smash{\mathop{\hbox to 30pt{\rightarrowfill}}\limits_{\scr #1}}}
\def\lrtu#1{\smash{\mathop{\hbox to 30pt{\rightarrowfill}}\limits^{\scr #1}}}

\def\que{\qu &\qu\rto \qu &\qu }
\def\fim{\quad{\vrule height6pt width6pt depth0pt}}
\def\dover{\over\displaystyle }
\def\tover{\over\textstyle }

\def\dov#1#2{{\displaystyle{#1}\over\displaystyle{#2}}}
\def\ex#1{\;\hbox{\rbig e}^{\scr #1}}
\def\ext#1{\;\hbox{\med e}^{\scr #1}}
\def\xvec(#1,#2,#3){
 \left(\begin{array}{c} {#1}\cr{#2}\cr{#3}\cr\end{array}\right) }

 \def\?#1{{\cal #1}}            
 \def\({\left }                 
 \def\){\right }                
 \def\[{\langle }
 \def\]{\rangle }

\def\z#1#2{\ifcase#1 {\overline {#2}} \or                   
{\null\ifmmode{\underline #2}\else{\underbar #2}\fi} \or    
{\if #2i {\hat\imath } \else\if #2j {\hat\jmath }           
\else {\hat {#2}} \fi\fi} \or
{\if #2i {\vec\imath} \else\if #2j {\vec\jmath}             
\else {\vec #2} \fi\fi} \or
{\if #2i {\tilde\imath} \else\if #2j {\tilde\jmath}         
\else {\tilde #2} \fi\fi} \or
{{\text{\b{$#2$}}}} \or   
{\!^#2} \or  
{\!_#2} \or  
{{\text{*}}#2} \or        
{{\bf #2}} \fi}                                           

\def\ddel{
\mathrel{\mathop{\nabla}\limits^{\leftharpoondown\kern-8pt\rightharpoonup}}
     \mathrel{\vphantom{\nabla}}
         }
\def\today{{\noindent \ifcase\month\or
January\or February\or March\or April\or May\or June\or July\or
August\or September\or October\or November\or December\fi
\space\number\day, \number\year}}
\def\del{\partial}

\def\sqr#1#2{{\vcenter{\vbox{\hrule height.#2pt               %
\hbox{\vrule width.#2pt height#1pt \kern#1pt
\vrule width.#2pt}
\hrule height.#2pt}}}}
\def\squ{\mathchoice\sqr{7}{6}\sqr{5.5}{5}\sqr{2.1}{3}\sqr{1.5}{3}} 
\def\dif#1{\,d\kern-1.3pt #1}                                      
\def\diff#1{\, d\kern-6truept\lower.2em\hbox{${^-}$}\kern-1.4pt #1}

 \def\za{{{}\ifmmode\alpha\else$\alpha$ \fi}}
 \def\zb{{{}\ifmmode\beta\else$\beta$ \fi}}
 \def\zc{{{}\ifmmode\psi\else$\psi$ \fi}}
 \def\zC{{{}\ifmmode\Psi\else$\Psi$ \fi}}
 \def\zd{{{}\ifmmode\delta\else$\delta$ \fi}}
 \def\zD{{{}\ifmmode\Delta\else$\Delta$ \fi}}
 \def\ze{{{}\ifmmode\epsilon\else$\epsilon$ \fi}}
 \def\zf{{{}\ifmmode\phi\else$\phi$ \fi}}
 \def\zF{{{}\ifmmode\Phi\else$\Phi$ \fi}}
 \def\zg{{{}\ifmmode\gamma\else$\gamma$ \fi}}
 \def\zG{{{}\ifmmode\Gamma\else$\Gamma$ \fi}}
 \def\zh{{{}\ifmmode\eta\else$\eta$ \fi}}
 \def\zi{{{}\ifmmode\iota\else$\iota$ \fi}}
 \def\zI{{{}\ifmmode\infty\else$\infty$ \fi}}        %
 \def\zk{{{}\ifmmode\kappa\else$\kappa$ \fi}}        %
 \def\zl{{{}{}\ifmmode\lambda\else$\lambda$ \fi}}
 \def\zL{{{}\ifmmode\Lambda\else$\Lambda$ \fi}}
 \def\zm{{{}\ifmmode\mu\else$\mu$ \fi}}
 \def\zn{{{}\ifmmode\nu\else$\nu$ \fi}}
 \def\zN{{{}\ifmmode\emptyset\else$\emptyset$ \fi}}  %
 \def\zp{{{}\ifmmode\pi\else$\pi$ \fi}}
 \def\zP{{{}\ifmmode\Pi\else$\Pi$ \fi}}
 \def\zchi{{{}\ifmmode\chi\else$\chi$ \fi}}
 \def\zvar{{{}\ifmmode\varphi\else$\varphi$ \fi}}
 \def\zr{{{}\ifmmode\rho\else$\rho$ \fi}}
 \def\zs{{{}\ifmmode\sigma\else$\sigma$ \fi}}
 \def\zS{{{}\ifmmode\Sigma\else$\Sigma$ \fi}}
 \def\zt{{{}\ifmmode\tau\else$\tau$ \fi}}
 \def\zu{{{}\ifmmode\upsilon\else$\upsilon$ \fi}}
 \def\zU{{{}\ifmmode\Upsilon\else$\Upsilon$ \fi}}
 \def\zv{\partial}                                   %
 \def\zV{\nabla}                                     %
 \def\zw{{{}\ifmmode\omega\else$\omega$ \fi}}
 \def\zW{{{}\ifmmode\Omega\else$\Omega$ \fi}}
 \def\zx{{{}\ifmmode\xi\else$\xi$ \fi}}
 \def\zX{{{}\ifmmode\Xi\else$\Xi$ \fi}}
 \def\zy{{{}\ifmmode\theta\else$\theta$ \fi}}
 \def\zY{{{}\ifmmode\Theta\else$\Theta$ \fi}}             
 \def\zz{{{}\ifmmode\zeta\else$\zeta$ \fi}}               
 \def\zZ#1{\ifcase#1 {}\or \displaystyle \or \textstyle   
        \or \scriptstyle \or \scriptscriptstyle \fi}      

 \def\scirc{\raise-.3em\hbox{\char'27}}
 \def\comp{\raise-.3em\hbox{\char'27}}
 \def\nill{\hbox{\rm\char'37}}
 \def\half{{1\over2}}
 
 \def\fourth{{1\over4}}

 \def\eighth{{1\over8}}

 \def\ahalf#1{{#1\over2}}

\newtheorem{theorem}{Theorem}
\newtheorem{definition}{Definition}
\newtheorem{casex}{Case}
\newtheorem{lemma}{Lemma}
\newtheorem{claim}{Claim}
\def\scri{{\cal J}}
\def\GG{G(\z9x_o,\z9x)}
\def\GR{G(\z9R)}
\def\dx0{|\z9x_o-\z9x|}
\def\psx{\varphi(\z9x)}
\def\psr{\varphi(\z9{x_o\!+\!R})}
\def\rcos{(R\cos\!\zy\!+\!x_o)}
\def\rcosa{[\rcos^2\!+\!a^2]}
\def\rt{(R\!-\!t_o)}
\def\sinq{\sin^2\!\zy}
\def\cosn{(x_o\cos\!\zy\!+\!t_o)}
\def\cost{(t_o\cos\!\zy\!+\!x_o)}
\def\cosy{\cos\!\zy}
\def\sinsin{\sin\!\zf\sin\!\zy}
\def\cossin{\cos\!\zf\sin\!\zy}
\def\cosq{\cos^2\!\zy}
\def\zcs{\biggl({\dis\zy\dover K_{m'}}\biggr)^{\ahalf3}}
\def\zcst{\biggl({\dis\zy\dover K_{m'}}\biggr)^{2}}

\def\A{A(\zy)}

\def\dom{d\zW}

\def\opi{{\dis1\dover 4\pi}}

\def\tht{{\theta\over 2}}
\def\be{{\bar e }}
\def\mbar{{{}\ifmmode\widetilde{\hbox{\rm M}}
\else${\widetilde{\hbox{\rm M}}}$\fi }}
\def\sbar{{{}\ifmmode\widetilde{\hbox{\rm S}}
\else${\widetilde{\hbox{\rm S}}}$\fi}}
\def\xbar{{{}\ifmmode\widetilde{\bf x}\else${\widetilde{\bf x}}$\ \fi }}
\def\ybar{{{}\ifmmode\widetilde{y}\else${\widetilde{y}}$\ \fi }}
\def\x{{{}\ifmmode{\bf x}\else${{\bf x}\ }$\fi}}
\def\M{{}\ifmmode{\hbox{\rm M}}\else{$\hbox{\rm M}$}\fi}
\def\H{{}\ifmmode{\hbox{\rm H}}\else{$\hbox{\rm H}$}\fi}
\def\B{{}\ifmmode{\hbox{\rm B}}\else{$\hbox{\rm B}$}\fi}
\def\X{{}\ifmmode{\hbox{\rm X}}\else{$\hbox{\rm X}$}\fi}
\def\G{{}\ifmmode{\hbox{\rm G}}\else{$\hbox{\rm G}$}\fi}
\def\vbar{{{}\ifmmode\widetilde{V}\else${\widetilde{V}}$\ \fi }}

\begin{document}
\preprint{ gr-qc/9507020 \quad\today}
\title{
\Large Cauchy Problem for Gott Spacetime
}
\bigskip
\author{Philip A. Carinhas\footnote{
Current address: \vskip-4pt
Departamento de Fisica Teorica,
Universitat de Valencia,
46100 Burjassot (Valencia) Spain
\vskip-4pt
(carinhas@kerr.ific.uv.es)} }
\address{
Department of Physics \\
University of Wisconsin--Milwaukee \\
Milwaukee, Wisconsin 53201
}
\maketitle
\smallskip
Gott recently has constructed a spacetime modeled by two infinitely long,
parallel cosmic strings which pass and gravitationally interact with each
other. For large enough velocity, the spacetime will contain
closed timelike curves.
An explicit construction of the solution for a scalar field is
presented in detail and a proof for the existence of such a solution is
given for initial data satisfying conditions on an asymptotically null partial
Cauchy surface.

Solutions to smooth operators on the covering space are invariant under the
isometry are shown to be pull back of solutions of the associated operator
on the base space.

Projection maps and translation operators for the covering space are
developed for the spacetime, and explicit expressions for the projection
operator and the isometry group of the covering space are given.
It is shown that the Gott spacetime defined is a quotient space of
Minkowski space by the discrete isometry subgroup of self-equivalences of the
projection map.
\newpage
\tableofcontents
\section{Introduction}
\label{gintro}

 The question of the validity of spacetimes with closed timelike curves has
been studied for several decades by many authors. See \cite{fried92,thorne92}
for a the state of the art review. It is not a simple task to
determine whether closed timelike curves (CTC's) are allowed in nature.
It has been shown that solutions in some nonchronal spacetimes
are exist for data on certain initial value surfaces \cite{fm91b,fmn90}.


 Gott has obtained a family of exact solution to Einstein's equations
describing
a system of two straight, infinitely long cosmic strings \cite{gott91}.
The resultant spacetime is flat except at conical singularities
representing the string loci.
The important feature of this spacetime is that, for sufficiently high
relative velocity, it contains CTC's \cite{boul92}.
Its global structure has been studied, and the region
containing CTC's has been determined \cite{cutler92}.

The question of whether or not a solution to fields that exist in the Gott
spacetimes is not in general known. The purpose of this work is to show that
for the scalar wave equation on a particular Gott spacetime,
asymptotically well behaved data
gives a solution everywhere except possibly at the string loci.

 The organization of the work is as follows. In \ref{geom}, we
outline the Gott two-string spacetime, (M,g).
In section \ref{cintro}, we discuss the covering space for M and how it
relates to the general solution for the massless wave equation and the
initial data surface.
Section \ref{proof} gives a formal proof that the Gott spacetime is covered
by Minkowski space.
Sections \ref{symmetry} and \ref{group} detail some of the
important symmetries of of the spacetime.
Finally, section \ref{intsoln} gives a general form for the scalar field in
terms of angular quadratures with the theorem
and proof that the integrals obtained are indeed defined and well behaved.
We discuss miscellaneous details in the appendices.
\newpage

\section{The Gott Spacetime}
\label{geom}

 In $2+1$ dimensions, the general vacuum solution to Einstein's equation
with a point mass is flat except for a conical singularities at the
location of the masses. In $3+1$ dimensions,
there are corresponding solutions with singularities on the worldsheets of
the strings. The mass (mass per length $\zm$ in the $3+1$ case)
is proportional to the deficit angle of the conical singularity
\cite{gott91}.

\firstfigfalse
\begin{figure}
\vbox{
\hfil
\epsfig{file=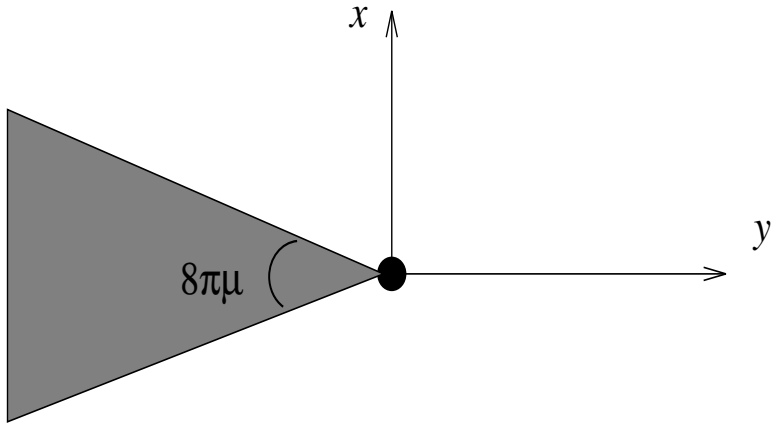,height=2.2truein,width=4.0in}
\hfill
}
\bigskip
\caption{The spacetime wedge cut out by a cosmic string. The deficit angle is
given by $8\pi\zm$, where \zm is the mass per unit length of the string. }
\label{wedge}
\end{figure}
\vskip.2truein

Two relatively moving strings, with their corresponding deficit angles,
will produce close timelike curves. If the sum of the
deficit angles is less than or equal to $\pi$, the resultant spacetime will
be open in the topological sense \cite{djt84}. The case where the deficit
angles are equal and sum to $\pi$ is considered in the work.
This will be essential in order that Minkowski space be a covering space
of the physical space.

Let the two strings parallel to the $z$ axis be located at
$x=0,y=\pm Y_0$ at $t= 0$.
In a static situation, we identify the $-x$ and $x$ components of each string.
The spacetime can thus be viewed as a $3+1$ slab with $-Y_0\leq y\leq Y_0$
that has been identified in the prior fashion.
\newpage

\begin{figure}
\vbox{\hfil
\epsfig{file=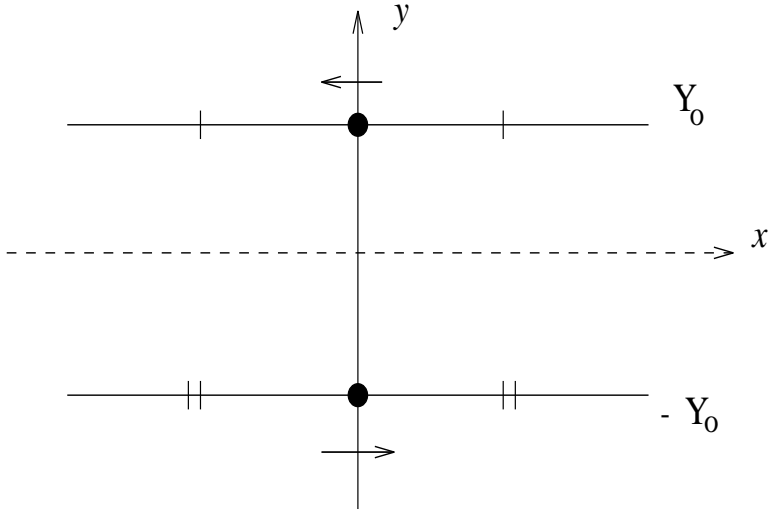,height=2.7truein,width=4.0in}
\hfill
}
\bigskip
\caption{The Gott spacetime as viewed along the string axis. The top string is
moving to the left and the bottom string moves to the right. The hash
marks indicate identification along the lines $y=\pm Y_0$.}
\label{gott2}
\end{figure}

In the dynamic case, the two strings are moving relative to one another in
$y$ as in figure \ref{gott2} and the identifications are Lorentz
transformations of the static case.

\bigskip

 Define null coordinates $x_\pm = x\pm t$ on Minkowski spacetime.
It will be shown that the Lorentz boosted identifications for the 2 string
system are

\beg \begin{array}{c}
x^\pm \to e^{\mp\za}x^\mp \quad  (y=+Y_0)             \cr
x^\pm \to e^{\pm\za}x^\mp \quad  (y=-Y_0) \label{id0} \cr
A_+: x^\pm \doteq -e^{\mp 2\za}x^\mp \quad  (y=+Y_0)             \cr
A_-: x^\pm \doteq -e^{\pm 2\za}x^\mp \quad  (y=-Y_0) \label{id1} \cr
\end{array} \eb

\<where $A_\pm$ are the identification maps at $\pm Y_0$.
\section{Covering Space for Gott Spacetime}
\label{cintro}

The proceeding section suggests that Gott Spacetime can be covered by
Minkowski spacetime. This will be demonstrated directly in section
\ref{projn}. Define the {\it unit cell} $D_o$ as the spacetime slab
bounded by the planes $y=\pm Y_0$. Similarly, define the {\it n-cell}
$D_n$ as

\beg D_n=\biggl\{ \xbar\in\mbar \big| (2n-1)Y_0<y< (2n+1)Y_0 \biggr\}
\label{dcell}.\eb

Gott spacetime can be characterized by the space $D_o$ with the boundaries
$y=\pm Y_0$ identified by the relations \ref{id1}.
We proceed to develop the explicit covering maps and other terminology
needed.
\subsection{Projection Maps for Stationary Strings}
\label{projn}

In this section, the projection map for the covering space is developed
by following a trajectory in Minkowski spacetime. The static case is
considered first, and afterwards, the dynamic case which contains
CTC's is discussed. The two spaces are topologically equivalent.

 The covering map for the stationary case follow from the fact that when one
crosses any boundary $y=(2n+1)Y_0$, the identifications \eqn{id1} reflect
through an angle of $\pi$ about the point $x=0$.

\begin{figure}
\vbox{\hfil \epsfig{file=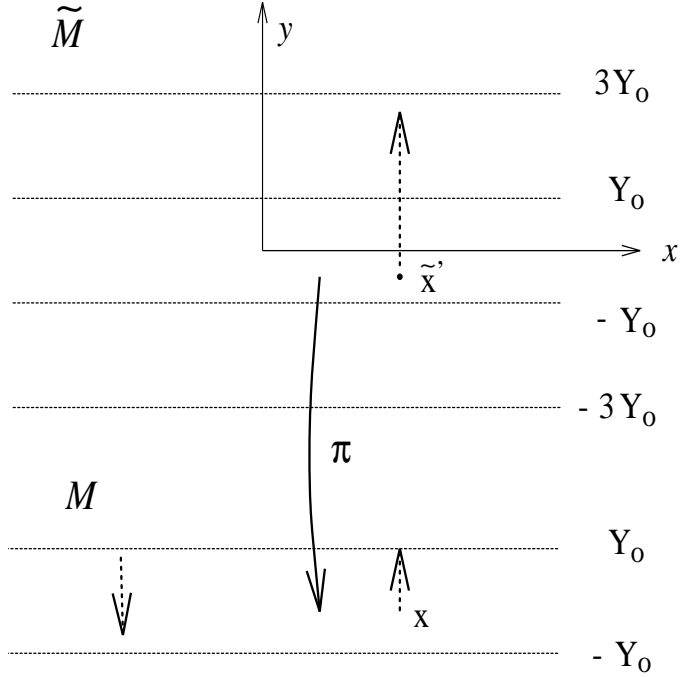,height=3.5truein,width=3.5in} \hfill }
\bigskip
\caption{A path in the covering space to determine the projection map $\pi$.
The dotted vertical line in \mbar\ represents the path in the covering space
that become the vertical paths in \M.}
\label{path}
\end{figure}

We can deduce the projection maps in the static case
by simply following a trajectory that moves vertically up in the $y$
direction as in figure \ref{path}.

{ \parindent=.1truein
\<1. For $ -Y_0 \leq y \leq Y_0$ $\pi$ is the identity:
\beg \pi(x,y,z)=(x,y,z) \eb

\<2. For $ Y_0 \leq y \leq 3Y_0$:
$x \rto -x \;;\; y \rto Y_0-\zD\equiv Y_0-(y-Y_0) = 2Y_0-y$
where $\zD$ is defined as the extra difference between the nearest line
$y=(2n+1)Y_0$ and the point in the path. Hence,

\beg \pi(x,y,z)=(-x,2Y_0-y,z) \eb

\<3. For $ 3Y_0 \leq y \leq 5Y_0$, $ x \rto x \;;\;
 y \rto -Y_0+\zD = -Y_0+(y-3Y_0)=y-4Y_0 $,

\<and so

\beg\pi(x,y,z)=(x,y-4Y_0,z).\eb

\<One can generalize the maps in the expression:

For $ (2n-1)Y_0 \leq y \leq (2n+1)Y_0$ \,,

\beg
\pi(x,y,z) = ((-1)^nx,(-1)^n(y-2nY_0),z)\;,\;\qu\forall x\in D_n \eb
}

\subsection{Projection Maps for Boosted Strings}

 The null $\z4x$ coordinates in the stationary frame defined by

\beg \z4 x^+ = \z4 x+\z4 t\qu;\qu \z4 x^- = \z4 x-\z4 t \eb

\<are used. Under a the lorentz transformation for the lower string at
$y=-Y_0$ moving in the $+x$ direction,

\beg x^\pm = e^{\pm\za} \z4 x^\pm\qu;\qu y = \z4 y - Y_0 \label{boost1} \eb

\<where the $\z4 x$ coordinates are stationary w.r.t. the point mass (string).
We identify antipodal (in $\z4 x$) points at $y = -Y_0$ using $A_-$ we obtain

\beg
A_-\z4 x=-\z4 x\imp A_-e^{-\za} x^+=-e^{\za}x^-\imp A_-x^\pm=-e^{2\za}x^\mp.
\eb

\<or equivalently,

\beg
A_-: x^\pm \longmapsto -e^{\pm 2\za}x^\mp \quad  (y=-Y_0) \label{id2}
\eb

\<Translating \eqn{id2} back to $(x,t)$ coordinates,

\beg
\begin{array}{ccc}
A_-: x\longmapsto -x \cosh 2\za + t \sinh 2\za \cr
A_-: t\longmapsto -x \sinh 2\za + t \cosh 2\za \cr
     \end{array}
      \qquad (y = -Y_0)
       \label{rec1}
        \eb

\<For the second point mass at $y=Y_0$ moving in the $+\z2x$ direction,

\beg
\begin{array}{lll}
A_+: x^\pm\!\longmapsto -e^{\mp 2\za}\z4x^\mp \quad \cr
A_+: x\;\longmapsto -x \cosh 2\za - t \sinh 2\za \cr
A_+: t\;\longmapsto +x \sinh 2\za + t \cosh 2\za \cr
\end{array} \qquad (y = Y_0)
\label{rec2}
\eb

We are now in a position to define our Gott spacetime.
Define a three dimensional spacelike slab, $W_0 = [-Y_0,Y_0] \times R^2$.
Let $S_0=R\times W_0 $ where the copy of $R$ corresponds to time $t$.
The Gott spacetime \M\ is defined by $S_0$ with identifications $A_\pm$
given by

\beg
A_\pm:(x^+,x^-,Y_0,z)\longmapsto (-e^{\mp 2\za}x^-,-e^{\pm 2\za}x^+,Y_0,z)
\quad @ \quad y = \pm Y_0
\label{ident0}
\eb

\<The action of the projections can be found by following the coordinates
in the base space sequentially through a path from $y=0$ to $y=2Y_0$ in
the covering space.

\beg
 \begin{array}{ccccc}
0  \que Y_0 \que 3Y_0  \cr
x^+\que -e^{-2\za}x^-\que -e^{+2\za}(-e^{+2\za}x^+)\cr
x^-\que -e^{+2\za}x^+\que -e^{-2\za}(-e^{-2\za}x^-)\cr
y    \que      y - 2Y_0     \que 4Y-0 - y
  \end{array}
  \eb

{ \parindent=.1truein
\<1. For $ -Y_0 \leq y \leq Y_0$ we again have the identity map;

\beg \pi(x^+,x^-,y)=(x^+,x^-,y).\eb

\<2. For $ Y_0 \leq y \leq 3Y_0$:

\beg \pi(x^+,x^-,y) = (-e^{-2\za}x^-,-e^{+2\za}x^+,2Y_0-y)
= K^{-1}P L^{-1}(x^+,x^-,y)\label{pione}\eb

\<3. For $ 3Y_0 \leq y \leq 5Y_0$:

\beg
\pi(x^\pm,y) = (e^{+4\za}x^+,e^{-4\za}x^-,y-4Y_0)
             =  KPLK^{-1}P L^{-1} (x^+,x^-,y)\label{pitwo} \eb
}

\<where $P$ is defined as the spatial parity operator, $K$ is the Lorentz
boost operator in $\hat x$ with velocity $\tanh\za$, and $L$ ($L^{-1}$) is
translation in $\hat y$ by $2Y_0$ ($-2Y_0$). The first two operators are
defined as

\beg
 \begin{array}{c}
 K(x^+,x^-,y)=(e^{-2\za}x^+,e^{+2\za}x^-,y)          \cr
 K^n(x^+,x^-,y)=(e^{-2n\za}x^+,e^{+2n\za}x^-,y)      \cr
 P (x^+,x^-,y)  =(-x^-,-x^+,-y)                      \cr
 P^n (x^+,x^-,y)  =(-1)^n(x^{(-)^n},x^{(-)^{n+1}},y) \cr
\end{array}
\eb

\< In the $(x^+, x^-,y)$ basis, $K$ and $P$ can be written as matrices:
\def\ph{\phantom{-}}

\beg
 K =
 \left[
  \begin{array}{ccc}
   e^{2\za}  & \  0 \     &\ 0\ \cr
  \  0\      & e^{-2\za}  &  0  \cr
     0       &    0       &  1  \cr
   \end{array} \right]
   \qquad
   P =
\left[
 \begin{array}{ccc}
\  0 \   &   -1  &\ 0\   \cr
   -1    &\ \ 0  &\ 0\   \cr
    0    &\ \ 0  & -1\   \cr
\end{array} \right]
\label{operator}
\eb

\<A few properties of $K^n$, $P^n$, and $L^n$ should be noted.

\beg
\begin{array}{c}
[K^n,L^m] = 0                                                          \cr
PL =L^{-1}P\quad;\quad P^n L^m=L^{(-1)^nm}=L^{\hbox{$s_{nm}$}}         \cr
K^n P =  P K^{-n} \quad;\quad K^n P^n = P^n K^{\hbox{$s_n$}}       \cr
 \end{array}
 \label{commute}
 \eb

\<where $s_n \equiv (-1)^nn$, $s_{nm} \equiv (-1)^nm$, and

\beg
 K^{\hbox{$s_n$}} =
 \left(
\begin{array}{ccc}
 e^{2{\hbox{$s_n$}}\za} &   \  0 \     &\ 0\ \cr
\  0\        & e^{-2\hbox{$s_n$}\za} &  0  \cr
   0         &      0       &  1  \cr
\end{array} \right)
\eb

{}From the above expressions, it follows that
\beg KPLK^{-1}P L^{-1} = K^2P^2 L^{-2}= K^2L^{-2}\eb

\<Thus the restriction $\pi_n$ of $\pi$ to $D_n$ where
$(2n-1)Y_0\leq y\leq (2n+1)Y_0$ \,, had the form

\beg \pi_n(x^+,x^-,y) =  K^{\hbox{$s_n$}} P^n L^{-n} (x^+,x^-,y), \eb

\<where we are regarding \M\ as the $O^{\hbox{th}}$-cell $D_o$ of Eq.
\eqn{dcell} withe identifications \eqn{id1}.

 \<Let's check against the result of Boulware \cite{boul92} for the set of
equivalent points $\pi^{-1}$.
 First, note that for $n=1$,

\beg
 \begin{array}{cl}
 \pi_1(x^+,x^-,y) &= K^{-1}P L (x^+, x^-,y) \cr
  &= \left(
       \begin{array}{ccc}
        e^{-2\za} &   \  0 \     &\ 0\ \cr
        \  0\     & e^{+2\za}    &  0  \cr
        0         &      0       &  1  \cr
    \end{array} \right)
\left(\begin{array}{c}
-x^-         \cr
-x^+         \cr
2Y_0- y      \cr
  \end{array} \right)
=\left(\begin{array}{c}
-e^{-2\za}x^-         \cr
-e^{+2\za}x^+         \cr
2Y_0- y               \cr
  \end{array} \right)                             \cr
  \end{array}
  \label{pi}
   \eb

\<This is equation (2.8) of reference \cite{boul92}. Next let $n=-1$

\beg
 \pi_{-1}\z9x   = KP L^{-1} \z9x
  = \left(
       \begin{array}{ccc}
        e^{+2\za} &   \  0 \     &\ 0\ \cr
        \  0\     & e^{-2\za}    &  0  \cr
        0         &      0       &  1  \cr
    \end{array} \right)
\left(\begin{array}{c}
 -x^-         \cr
-x^+         \cr
{\scr -y-2Y_0}\cr
  \end{array} \right)
=\left(\begin{array}{c}
-e^{+2\za}x^- \cr
-e^{-2\za}x^+ \cr
{\scr -y-2Y_0}\cr
  \end{array} \right)
   \eb

\<which is equation (2.6) of \cite{boul92}.
Note that in the static case ($e^{\pm2n\za}\to 1$), $\pi$ can be rewritten as

\beg \pi_n(x^+,x^-,y) = P^n L^{-n} (x^+, x^-,y).\eb

\<which is the correct limiting form.
\section{Theorem for the Covering Space}
\label{proof}

\begin{theorem} Let \mbar\ be Minkowski spacetime with the string
worldsheets removed, and let $\zp:\mbar\rto\M$ be
defined as above. Then \mbar\ is a covering space of \M\ and
$\zp:\mbar\rto \M$ is the projection map.
\end{theorem}

\<We prove the theorem with the following

\begin{claim} $\pi$ is a local homeomorphism in the static case. \end{claim}

\<Proof:

  Take a point $\be= (\bar e_x,\bar e_y)$ that lies interior to all lines
$y = (2n+1)Y_0$ such that $\zp({\bar e})= e$. Then there is an open nbhd
$U_\be$ such that

\beg
U_\be
=\{(x,y)\in\mbar\Bigm|
  |(x,y)-(\be_x,\be_y)|<\min(|(x,y)-(x,(2n+1)Y_0)|) \forall n\}
\eb

\<Then $\zp(U_{\bar e})$ is mapped homeomorphically onto an open set
$V_e$ since $\pi$ is a polynomial map of degree one, which is
continuous and invertible.

Now take a point $\be =(x,(2n_0+1)Y_0)$ for some $n_0$. Let $U_{\be}$ be
defined as
\beg
U_{\be} = \{(x,y)\in \mbar \Bigm| |(x,y)-(e_x,e_y)| < \ze \}
\eb

\<where

\beg
\ze\equiv\min\Bigl(|\be-(0,(2n_0+1)Y_0)|,Y_0\Bigr).
\eb

\<With this choice of $\ze$ we ensure that we neither intersect the string nor
any other line $y=(2n+1)Y_0\quad n\neq n_0$. Let

\beg
 \begin{array}{llc}
U_0 &\equiv &U_\be \bigcap \{y =  (2n_0+1)Y_0\}\cr
U_+ &\equiv &U_\be \bigcap \{y\geq(2n_0+1)Y_0\}\cr
U_- &\equiv &U_\be \bigcap \{y\leq(2n_0+1)Y_0\}.\cr
 \end{array}
  \eb

\<Then points of $U_0$ are mapped to two different regions of
  $R \times [-Y_0,Y_0]$ of M:

\beg
 \begin{array}{ccccc}
 \zp_{n_0}\mid_{U_0}(x,y)
 &= &(-1)^{n_0}(x,y-2n_0Y_0) &=&(-1)^{n_0}(x,Y_0) \cr
 \zp_{n_0+1}\mid_{U_0}(x,y) &= &(-1)^{n_0+1}(x,y-2(n_0+1)Y_0) &
   =&(-1)^{n_0}(-x,Y_0), \cr
  \end{array}
    \eb

\<which are identified by the map $A_+$. So the interiors of $U_+$ and $U_-$
are
mapped homeomorphically onto M as above, and agree on the boundary $U_0$.
Thus, the whole of $U_\be$ is mapped homeomorphically onto M by $\pi$.\fim

\begin{claim} $\pi$ is a local homeomorphism in the dynamic case. \end{claim}
\<Proof:

 Denote the static projection by $\pi_s$ and the dynamic projection by $\pi$.
Let $K^{\hbox{$s_n$}}$ denote the appropriate Lorentz transformation which is
applied in the projection operator $\pi$. Then at each point in the covering
space \mbar, the projection map can be expressed as

\beg\zp = K^{\hbox{$s_n$}}\comp \pi_s\eb

\<Noting that the Lorentz transformation is a homeomorphism and that
composition of homeomorphisms are homeomorphisms, $\pi$ is a homeomorphism.\fim

\begin{claim} Every point of $e \in \M$ has a neighborhood $V_x$ which is
mapped to a disjoint union of open subsets of $\mbar$ by $\pi^{-1}$.
\end{claim}

\<Proof:

Take a point $e=(t,e_x,e_y)=(e^+,e^-,e_y)$ in M that does not lie on
$y=\pm Y_0$.

Then there is an open neighborhood $V_e$ defined by

\beg
V_e = \{x = (x^+,x^-,y)\in \M \Bigm| |(x^+,x^-,y)-(e^+,e^-,e_y)| < \ze \}
\eb

\<where \ze is chosen to be the smaller of the distances to either line
$y=\pm Y_0$ (this also guarantees that $e$ does not intersect any string
locus):

\beg
\ze = \min\Big(|(e^+,e^-,y)-(e^+,e^-,Y_0)|,|(e^+,e^-,y)-(e^+,e^-,-Y_0)|\Big).
\eb

\<Since it has been shown that \zp is a local homeomorphism, (\zp maps open
sets to open sets), we need only show that $\zp^{-1}(V_e)$ is a disjoint union
of sets in \mbar. Write $\zp^{-1}$ in a similar fashion to \eqn{pi};
 $ \pi^{-1}:\M\rto\mbar$ mapping $x$ to the $n^{th}$ level of \mbar\ where
$(2n-1)Y_0 \leq y \leq (2n+1)Y_0$ \, ,

\beg
 \begin{array}{lll}
\pi_n^{-1}(x^+,x^-,y)
    &=&L^n P^n K^{-\hbox{$s_n$}}(x^+, x^-, y )   \cr
    &=&K^{-n} P^n (x^+, x^-, y )+ (0,0,2nY_0)   \cr
 \end{array}
\label{pi_inv}
\eb

\<where we have used equation \eqn{commute}. For this choice of $e$,
$\pi_n^{-1}$ maps $V_e$ to each level of \mbar\ for
each $n$ while it does not affect the diameter of $U_e=\pi_n^{-1}(V_e)$ in $y$.
By construction, $\pi_n^{-1}(V_e)$ does not intersect any line
$y=(2n+1)Y_0$ for any $n$ and so all images of $\pi_n^{-1}(V_e)$ are
disjoint.

 Now take a point $e=(e^+,e^-,Y_0) \in\M$. Define $V_e$ by
\beg
V_e = \{x = (x^+,x^-,y)\in \M \Bigm| |(x^+,x^-,y)-(e^+,e^-,Y_0)| < \ze \}
\eb

\<where now, \ze is defined as $ \ze = \min\Big(Y_0, d(\z4 0,e)\Big)$,
where $\z4 0$ is the the locus of the (upper) string in its local coordinates.
This choice ensures that $V_e$ does not intersect the string on the line
$y=Y_0$ or the lower string at $y=-Y_0$.

 As in the static case, $V_0$ is the sum of 2 disjoint regions in
$R\times [-Y_0,Y_0]$ before the identifications are made on $y=\pm Y_0$;

\beg
 \begin{array}{lll}
V_{0} &\equiv  & \{(x^+,x^-,Y_0)\in V_e \}\cr
V_{e+} &\equiv & \{(x^+,x^-,y)\in V_e | e^{+\za}x^++e^{-\za}x^- > 0 \}\cr
V_{e-} &\equiv & \{(x^+,x^-,y)\in V_e | e^{+\za}x^++e^{-\za}x^- < 0 \}.\cr
\end{array}
 \label{sets}
  \eb

\<This division effectively separates the neighborhood $V_e$ into
2 components on either side of the string at $y=Y_0$.
(Note that $(e^{+\za}x^++e^{-\za}x^-=0;y=Y-0)$ describes the locus of string
on $y=Y_0$.) To see this, note that at $y=Y_0$,

\beg A_+: (x^+,x^-,Y_0)=(-e^{-2\za}x^-,-e^{+2\za}x^+,Y_0) \}, \label{ident1}\eb

\<so that

\beg A_+:(e^{+\za}x^++e^{-\za}x^-)=-(e^{+\za}x^++e^{-\za}x^-)\label{ident2} \eb

\<which is clearly less than zero if \quad $e^{+\za}x^++e^{-\za}x^- > 0$,
as in the second of equations \eqn{sets}. Finally note that distances are
preserved under the identification map $A_+$, which ensures that
$V_{e+}\cap V_{e-}=\emptyset$.

\beg d(x-e)= x^+x^- +Y_0Y_0
=(-e^{-\za}x^-)(-e^{+\za}x^+)+Y_0Y_0 = d(A_+:x-A_+:e)
 \eb
Now, $\pi_n^{-1}(V_e) = \pi_n^{-1}(V_{e+})\cup \pi_{n+1}^{-1}(V_{e-})$, where

\beg
 \begin{array}{l}
\pi_n^{-1}(V_{e})\equiv\{(x^+,x^-,y)\in D_n\,|\,\pi(x^+,x^-,Y_0)\in V_{e} \}\cr
\pi_n^{-1}(V_{e+})\equiv\{(x^+,x^-,y)\in D_n\,|\,\pi(x^+,x^-,Y_0)\in
V_{e+}\}\cr
\pi_{n+1}^{-1}(V_{e-})\equiv\{(x^+,x^-,y)\in D_{n+1}
|\,\pi(x^+,x^-,Y_0)\in V_{e+}\}\cr
 \end{array}
  \label{pi_inv2}
   \eb

\<Assume for the moment that $n$ is even
(the argument is the same for odd $n$).
Then $\pi_n^{-1}(V_{e+})$ is mapped onto the strip
$(2n-1)Y_0 \leq y \leq (2n+1)Y_0$, and $\pi_{n+1}^{-1}(V_{e-})$
is mapped onto the strip $(2n+1)Y_0 \leq y \leq (2n+3)Y_0$,
and they agree on $V_0$. Note that for even $n$, $s_n = n$ and $P^n = 1$.

\beg
  \pi_n^{-1}\z9x= L^{n}P^n K^{-\hbox{$s_n$}}\z9x =
  L^{n} \left( \begin{array}{ccc}
 e^{-2n\za} &   \  0 \     &\ 0\ \cr
  \  0\     & e^{2n\za}    &  0  \cr
     0      &      0       &  1  \cr
  \end{array} \right)
  \left(\begin{array}{c}
           x^+           \cr
           x^-           \cr
           Y_0           \cr
\end{array} \right)
=   \left(\begin{array}{c}
           e^{-2n\za}x^+ \cr
           e^{2n\za} x^- \cr
       {\scr (2n+1)Y_0 } \cr
\end{array} \right)
\eb

\beg
 \begin{array}{c}
  \pi_{n+1}^{-1}A_+(x^+,x^-,Y_0)
 = L^{n+1}P K^{-\hbox{$s_{n+1}$}}(-e^{-2\za}x^-,-e^{2\za}x^+,Y_0) \cr
\noalign{\vskip8pt}
 = L^{n+1}P^n \left( \begin{array}{ccc}
  e^{-2{\hbox{$s_{n+1}$}}\za} &      0        &  0  \cr
     0      &\!\! e^{2{\hbox{$s_{n+1}$}}\za}\ &  0  \cr
     0      &                        0        &  1  \cr
    \end{array} \right)
\left(\begin{array}{c}
-e^{-2\za}x^- \cr
-e^{ 2\za}x^+ \cr
 Y_0          \cr
\end{array} \right)                               \cr
\noalign{\vskip10pt}
    =  L^{n+1}P
    \left(\begin{array}{c}
          -e^{2n\za}x^-  \cr
          -e^{-2n\za}x^+ \cr
           Y_0           \cr
          \end{array} \right)
    =  L^{n+1}
    \left(\begin{array}{c}
           e^{-2n\za}x^+ \cr
           e^{2n\za}x^-  \cr
           -Y_0          \cr
          \end{array} \right)                     \cr
\noalign{\vskip8pt}
=   \left(\begin{array}{c}
           e^{-2n\za}x^+ \cr
           e^{2n\za} x^- \cr
             (2n+1)Y_0   \cr
\end{array} \right) = \pi_n^{-1}(x^+,x^-,Y_0)
\end{array}
\eb

\<where the identification $A_+$ is defined in \eqn{ident1}. Thus $V_e$ is
mapped to $(2n-1)Y_0-\ze \leq y \leq (2n+1)Y_0+\ze$ by $\zp^{-1}$ and does not
intersect any other image of $\pi^{-1}(V_e)$ which is $2Y_0 - 2\ze>0$ away.
Thus $\pi^{-1}$ maps $(V_e)$ to disjoint open sets in \mbar\ and \zp is
indeed a the projection map.\fim
\section{Spacetime Symmetries }
\label{symmetry}
We take the covering space to be Minkowski space, and $\pi$ be the
projection map from \mbar\ to \M.

A solution \zc of the wave equation on the covering space is the pullback of
a solution on the physical space by the projection map, provided that the
solution on the cover is invariant under the isometry group $H$ that
relates equivalent points of $\pi^{-1}_m$.

Let the elements of the group $H$ be $h_n$
taking a point $x\in\mbar$ n cells above it.
The above condition implies that $h_n\zc = \zc$ for all $n$.
It is therefore important to study the projection mapping $\pi_n$ and
the isometry group $\{h_n\}$.

\begin{figure}
\vbox{
\hbox{\hskip1.0truein
\epsfig{file=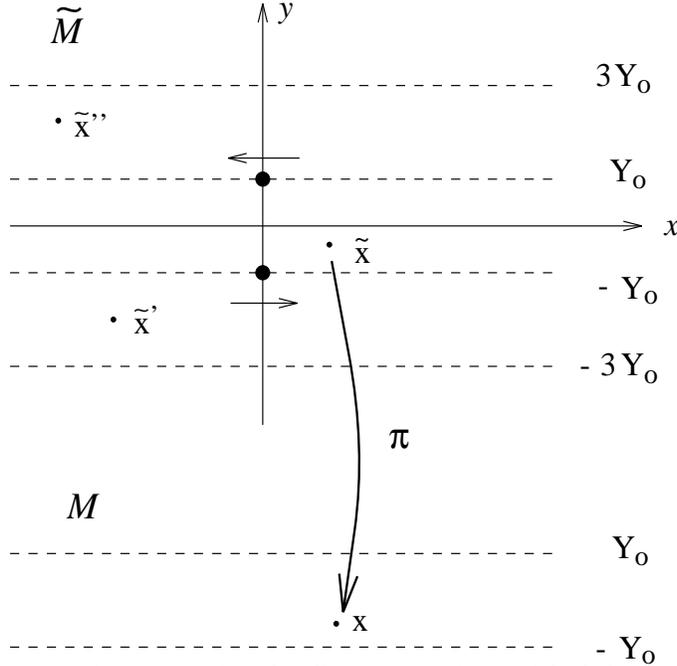,height=3.5truein,width=3.5truein}
\hfil
}}
\caption{The covering space for the Gott spacetime with deficit angel equal to
$2\pi$. The physical (base) space $M$ is defined between $-Y_0\leq y\leq Y_0$.
The identifications $A_\pm$ occur at $\pm Y_0$ in $\sigma$.
The points $\tilde x_o$, $\tilde x_o'$, and  $\tilde x_o''$, which
differ by elements of $H$, are all mapped to $x_o$ by $\pi$. }
\label{gott1}
\end{figure}
\medskip

We now wish to establish that a solution \zc exists that is invariant under
$H$ and solves the scalar wave equation,

\beg \squ\zc =(\del_x^2 +\del_y^2 +\del_z^2 -\del_t^2 )\phi = 0. \eb

\<The general solution on \mbar\ can be written as an integral of the
retarded Green's function $\GG$ over the initial data

\beg \zS=\{(t,x,y,z)|\;t+(x^2+a^2)^\half =0 \}.\eb

\<The integral has the form

\beg
\begin{array}{ll}
\zc(\z9x_o)
&=\opi\intl_{\zS} \GG\ddel_\za \psx dV^\za \cr
&=\opi \intl_{\zS} \{ \GG\zv_\za \psx - \psx \zv_\za\GG\}dV^\za.
\end{array}
\eb

\<Here $dV^\za=n^\za dS$, where $dS$ is the invariant 3-surface element
\<and $n^\za$ is the unit normal to $\zS$.

\<The point $\z9x_o$ lies within the unit cell $D_o$. The initial data
surface restricted to $D_o$, labeled $\zs$ is defined by

\beg \zs = \zS \cap \{-Y_0 \leq y \leq Y_0 \}. \eb

\<Then $\zS$ can be rewritten as

\beg \zS = \cup_n h_n \zs . \eb

\<We may rewrite our solution as

\beg
\begin{array}{l}
\zc(\z9x_o)
=\opi\intl_{\cup_n h_n\zs} \GG\ddel_\za \psx dV^\za
=\opi\Sum_n\intl_{h_n\zs} \GG\ddel_\za \psx dV^\za \cr
=\opi\Sum_n\intl_{\zs} h_n^{-1}\{\GG\ddel_\za\psx dV^\za \}
=\opi\Sum_n\intl_{\zs}G(\z9x_o,h_n^{-1}(\z9x))\ddel_\za\psx dV^\za\cr
=\opi\Sum_n\intl_{\zs}G(h_n\z9x_o,\z9x)\ddel_\za\psx dV^\za\cr
\end{array}
\eb

\<Where we used the fact that $\psx$ and $\ddel\,\;$ are invariant under
the isometry $h_n$ and that

\beg
G(h_n\z9x,h_n\z9y)=G(h_n|\z9x\!-\!\z9y|)=G(|\z9x\!-\!\z9y|)=G(\z9x,\z9y)
.\label{partial}
\eb

\<It is clear now that $\zc(\z9x_o)$ will be $h$ invariant since the sum

\beg
\begin{array}{ll}
h_m\zc(\z9x_o)
&\qu=\qu\opi\Sum_nh_m\intl_{\zs}G(h_n\z9x_o,\z9x)\ddel_\za\psx dV^\za\qu\cr
&\qu=\qu\opi\Sum_n\intl_{\zs}G(h_{m+n}\z9x_o,\z9x)\ddel_\za\psx dV^\za\qu\cr
&\qu=\qu\opi\Sum_{n'}\intl_{\zs}G(h_{n'}\z9x_o,\z9x)\ddel_\za\psx dV^\za\qu.\cr
\end{array}
\eb

In order to simplify the calculations, we consider data on a cell
that is twice the original width of the unit cell called the two-cell.
The data is still only unique on the unit cell, but
the form of the translation operator $h_n$ is simplified by restricting it
to only even $n=2m$. The translation operator will take the form,

\beg h_n(x^+,x^-,y,z)=(e^{4m\za}x^+,e^{-4m\za}x^-,y-4mY_0,z).\label{trans}\eb

\<This relation will be explicitly derived in appendix \ref{isogroup}.
Denote the solution in the two-cell by $\zc_\zs$ and similarly for the
coordinates. When the integrand of the solution lies inside the $m^{th}$
two-cell, the value of the initial data is as follows:

\beg
\zvar_\zs(x^+,x^-,y,z) =
\zvar(e^{4m\za}x^+,e^{-4m\za}x^-,y-4mY_0,z)
\eb

\<where $\quad (4m-1)Y_0\leq y \leq (4m+3)Y_0 \;$.
Under this transformation, the Cartesian coordinates become

\beg
\begin{array}{l}
t_\zs= x \sinh(4m\za)+ t \cosh(4m\za)\cr
x_\zs= x \cosh(4m\za)+ t \sinh(4m\za)\cr
y_\zs= y - 4mY_0                     \cr
z_\zs= z                             \cr
\end{array}
\label{lorentz}
\eb

For the solutions on the covering space to give solutions in the physical
(base) space, we need the following lemma. For a general proof, see
section \ref{gappen}.

\begin{lemma}
Let \mbar\ be a covering space for $M=\mbar/H$ and $\tilde\zc$ be a solution
of $\squ\tilde\zc=0$ in \mbar\ invariant under $H$.
Then \hbox{$\tilde\zc = \zc\circ\ \pi$} for a solution satisfying $\squ\zc=0$
on $M$.
\end{lemma}

\<The proof follows from the fact that any $\tilde\zc$ invariant under $H$
has the form \hbox{$\tilde\zc = \zc\circ\ \pi$}, for a function $\zc$ on \M.

\section{Group Action on $\widetilde{\hbox{M}}$ }
\label{group}
  In this section we will develop the isometry group \H\ on the covering space.
Explicit formulae have already given for the elements of \H\ in \ref{cintro}.
Proceed to find the generators of $h$ in terms of the operators
$K$, $L$, and $P$ defined in \eqn{operator}. We can find the two generators
by consiering the projection maps given in \eqn{pione} and \eqn{pitwo}.

 The translation map is locally the inverse of the projection operator.
At the bottom identification $y=-Y-0$, $\pi^{-1}=(K^{-1} P L^{-1})^{-1}$.
{}From the commutation relations \eqn{commute}, we have that
$\pi^{-1}=\pi$. The generator $L^{-1}KP$ is its own inverse.

Define $h_\pm$ by

\beg h_-= L^{-1} K P \;;\;h_+ = L K^{-1} P.\eb

It is clear that since $K$ and $P$ are linear operators and $L$ is a
translation operator, that H defines an associative action on \mbar.
Also, from \eqn{commute}

\beg h_-^{-1}= h_- \;;\;h_+^{-1} = h_- .\eb

The relevant elements of $H$ are therefore of the form

\beg h_-h_+h_-h_+\cdots \;;\; h_+h_-h_+h_-\cdots \eb

\<without any repetitions.
The group is thus generated by the two elements $h_\pm$, for translation
$h_{2n}$ can be written as

\beg h_{2n}=(h_+ h_-)^n= L^{2n}K^{-2n}\label{transn}. \eb

 We now show that only even $n$ lead to closed timelike curves.
Consider whether 2 points in the covering space are timelike separated.
Those points, when projected down to the physical space, form CTC's.

To find out, take a point and act on it by $h_{n}$ and then take the
resulting difference and contract it with the metric. Assume that $X$
is located at the $l^th$ level of the covering space.
First take the case where we translate up an even number of levels $2n$
so that we may use the former formulae above. Then the translated point is

\beg (e^{-4n\za}X^+,e^{4n\za}X^-,Y+4nY_0)= L^{2n}K^{-2n}(X^+,X^-,Y) \eb

\<and the displacement vector is then

\beg
\begin{array}{ll}
\zD X &= (e^{-4n\za}X^+,e^{4n\za}X^-,Y+4nY_0)-(X^+,X^-,Y)      \cr
      & = \left[X^+(e^{-4n\za}-1),X^-(e^{4n\za}-1),4nY_0\right] \cr
\end{array}
\eb

\<We now wish to establish when this vector is timelike, ie.

\beg\begin{array}{ll}
g(\zD X,\zD X) =&  X^+X^- (2- e^{-4n\za}-e^{4n\za})+16n^2Y_0^2 \leq 0 \cr
\Rightarrow &2X^+X^- (1- \cosh 4n\za)  + 16n^2Y_0^2 \leq 0 \cr
\Rightarrow & X^+X^- \geq{\dis 8n^2Y_0^2\dover(\cosh 4n\za -1)}\cr
\end{array}\eb

\<Which defines the open region bound by the a hyperbolas $X^+X^-=a$
($a$ constant) since the left hand side is always a positive constant.
Notice in the limit of $\za\to 0$,
there are no timelike paths that join
identified points, but that for any nonzero value for \za, there are
paths at large distances from the origin.

Next take the same point and translate an by odd number $2n+1$.
Then the prior analysis gives that

\beg
\begin{array}{ccr}
h_{2n+1} X &= (-e^{-2(2n+1)\za}X^-,-e^{2(2n+1)\za}X^+,-y-2(2n+1)Y_0)&,\cr
g(\zD X,\zD X)
&= 2X^+X^-+e^{-2(2n+1)\za}X^-X^-+e^{2(2n+1)\za}X^+X^+&\cr
&+4[y-(2n+1)Y_0]^2\leq 0&
\end{array}
\label{inequal} \eb

\<This restriction is impossible to achieve since
if we view the above equation as a quadratic equation in $X^+$,
the solution requires that the radical term in the quadratic solution

\beg
2X^-X^- - 4 e^{(2n+1)\za}(e^{-(2n+1)\za}X^-X^- + 4[y-(2n+1)Y_0]^2)
\eb

\<be positive. But this is impossible since

\beg
\begin{array}{l}
2X^-X^- - 4 e^{(2n+1)\za}(e^{-(2n+1)\za}X^-X^- + 4[y-(2n+1)Y_0]^2) \cr
= 4(X^-)^2 - 4 (X^-)^2 -16e^{2(2n+1)\za} [y-(2n+1)Y_0]^2 \cr
= -16e^{2(2n+1)\za} [y-(2n+1)Y_0]^2 \cr.
\end{array}
\eb

\<which is always negative definite.
Furthermore, if the inequality is strict in equation \eqn{inequal},
then adding a positive constant to the expression and
solving again for the quadratic equation only makes the radical more negative,
and so there is no solution for equation \eqn{inequal}. There can thus be
no CTC's that are connecting an odd number of copies in the covering space.
We may therefore focus our attention on the even solutions.

\subsection{Gott Spacetime as a Quotient Space}
The Gott spacetime described above can be characterized as a quotient space
of $\mbar/H$. This is no surprise since Minkowski space is its cover and
$H$ comes from the identification maps. The following section defines
a quotient space and shows that this is equivalent to the GST.

Define $\S=\mbar/\H$. Recall the definition of a quotient space:

\begin{definition}
Given a manifold \X\ and a group \G\ that acts on \X\, the quotient space
\X/\G\ is defined as the space $\X/\!\!\sim$ , where $\sim$ is the equivalence
relation $x_1\sim x_2$ iff $x_1=g(x_2)$ for some $g\in\G$.
\end{definition}

\<Recall the definition of the space $D_n$:

\beg D_n=\biggl\{ \xbar\in\mbar \bigg| (2n-1)Y_0<y< (2n+1)Y_0 \biggr\}.\eb

\<Let $\pi:\mbar\longmapsto\S$ be the projection operator defined in
\ref{projn}.
First take $\xbar\in\S$ with $\x=\pi(\xbar)$ such that the $y$ coordinate
of \xbar does not lie in $(2m+1)Y_0$ for any $m$
(then there exists a unique $\xbar_o\in D_0$ such that $\pi(\xbar_o)=\x$).
There exists an $n$ such that for $(2m-1)Y_0<\z4y<(2m+1)Y_0$,
$h_n(\xbar)\in D_0$. Thus

\beg D_0\simeq\biggl\{\xbar\in\S\biggm|\pi^{-1}(\x)\cap(2n+1)Y_0=0\biggr\}.\eb

  Now take $\xbar\in\S$ such that $\pi^{-1}(x)$ intersects $(2n+1)Y_0$
for some $n$. Then there exists an $n$ such that $h_n(\ybar)=\ybar$, and
so there are two representatives of $\pi^{-1}(x)$ at $y=-Y_0$ for example.
Choosing a unique representative is equivalent to identifying the same
points by the identification maps $A_\pm$. Note that at the loci of the
strings, there is only one representative. This implies the following
identity:

\beg \overline{D}_0/A_\pm \equiv (\hbox{Closure of $D_0$})/A_\pm \eb

This is precisely our current definition of the Gott spacetime.

\section{Integral Form for the Solution}
\label{intsoln}

It is critical to develop a form for the solution that takes advantage of
the symmetry of our GST. Doing so will facilitate the analysis of the
solution. In the following sections, we will find conditions
on the data and expressions for the solution that will do exactly this.

\subsection{Behavior of the Initial Data}
\label{initial}

 We wish to establish the asymptotic form for the data on the Gott spacetime.
The asymptotics for the data on the covering space will be determined by the
pullback of the data $\pi^{-1}(\varphi)$ to Minkowski space and the action of
the translation operator $h_n$ of section \ref{symmetry}.
We assume the initial data $\zvar$ and $\zv_\za\zvar$ on $\zS$ will behave
asymptotically as a power law to first order in the $x$ and $z$ coordinates.
Asymptotic powers for the solution and its derivatives
will be found that make the energy condition finite in Minkowski space,
will suffice to give a well defined solution in the Gott Spacetime.

As before, express the solution to the initial value problem as

\beg
\zc(\z9x_o) =\opi\intl_\zS \GG\ddel_\za \psx dV^\za
\eb

\<where $M$ is Minkowski spacetime and $\GG$ is defined by

\beg \GG = {\displaystyle \zd(t_o-t-\dx0) \over\dis \dx0}.\eb

\<Assume the integrand asymptotically
behaves as a power law, $\dis 1\dover r^{1\sscr}$,

\beg
\begin{array}{l}
\psx\del_\zm \GG r^2 = O\!\left({\dis 1\dover r^{1}}\right)\cr
\noalign{\vskip10pt}
\GG\del_\zm\psx r^2  = O\!\left({\dis 1\dover r^{1}}\right).
\end{array}
\label{cond1}
\eb

\<On the other hand, demand that the total energy be finite,

\beg
\intl_\zS
\biggl\{
\zvar^2 - (\zv_t\zvar)^2 +({\dis \zV\zvar})^2
\biggr\}d^3\!V <  \infty.
\label{cond2}
\eb

\<Assuming the solution satisfies the conditions \eqn{cond1},
 \eqn{cond2}, the data must satisfy

\beg
\zvar = O(R^{-q})
\qu;\qu
\zv_\zm\zvar =O(R^{-p}),
\label{data0}
\eb

\<where $q=\ahalf3$ and $p=2$. In particular there exists constants $K$ and
$K_\zm$ such that

\beg
\zvar\leq KR^{-\ahalf3}\qu;\qu\zv_\zm\zvar\leq K_\zm R^{-2}.\label{data1}
\eb

It is understood that $R$ in this case can only extend out to arbitrarily
large values in the $x$ and $z$ directions. It should be noted that these
conditions are stricter than what is needed to make the solution converge.

\subsection{Reduction of the Solution to Quadratures}

Given the initial data surface above, the 3-metric restricted to $\zS$ is

\beg
 \begin{array}{cl}
 ds^2 =  -dt^2 + dx^2 + dy^2 +dz^2
 &= -\left({\dis x^2\dover x^2+a^2}\right) dx^2 + dx^2 + dy^2 +dz^2 \cr
\noalign{\vskip8pt}
= \left({\hbox{$ a^2$}\over\hbox{$x^2+a^2$}}\right)dx^2  + dy^2 +dz^2 &.\cr
\end{array}
\eb

\<where $\zS$ is given again by

\beg \zS=\{(t,x,y,z)|\;t+(x^2+a^2)^\half =0 \}.\eb

\vskip5pt
\<In matrix form,

\beg ^3g_{ij} = \left[
 \begin{array}{ccc}
{\dis a^2\dover x^2+a^2} &\; 0 &\qu 0\,  \cr
           0       &\; 1 &\qu 0  \cr
           0       &\; 0 &\qu 1  \cr
   \end{array}\label{2metric}
\right]
 \eb

\<The normal to $\zS$ is then

\beg
n^\za=
\left[ \begin{array}{c}
{-t/a } \cr
{-x/a}  \cr
{0 }    \cr
{0 }    \cr
\end{array}\right]
=
\left[ \begin{array}{c}
{\sqrt{\dis x^2+a^2}\over\dis a } \cr
{-x/a}  \cr
{\quad 0 \quad}    \cr
{\quad 0 \quad}    \cr
\end{array}\right]
\quad ;\quad
n_\za =
\left[ \begin{array}{c}
{ t/a } \cr
{-x/a}  \cr
{\quad 0 \quad}    \cr
{\quad 0 \quad}    \cr
\end{array}\right]
=
\left[\!
\begin{array}{c}
-{\sqrt{\dis x^2+a^2}\dover a } \cr
{-x/a}  \cr
{\quad 0 \quad}    \cr
{\quad 0 \quad}    \cr
\end{array}\right]
\eb

\<and we have

\beg
dV^0 = {\dis -t\dover (x^2+a^2)^\half }\; dx dy dz = dx dy dz
\quad ; \quad
dV^x = {\dis -x\dover (x^2+a^2)^\half }\; dx dy dz.
\eb

\< The solution takes the form

\beg
\begin{array}{l}
\zc(\z9x_o)\quad =\quad \cr
{\hbox{$1$}\over\hbox{$4\pi$}}
{\intl_{\zS}}
\{ \GG\zv_o \psx - \psx \zv_o \GG \} \; d^3\!x  \cr
\noalign{\vskip2pt}
-{\hbox{$1$}\over\hbox{$4\pi$}} {\intl_{\zS}}
\{ \GG\zv_1 \psx - \psx   \zv_1 \GG \}
\;{\hbox{$x$}\over\hbox{$(x^2+a^2)^\half$}}\;d^3\!x  \cr
\noalign{\vskip10pt}
\quad =\quad {\dis 1\over\dis 4\pi} {\intl_{\zS}}
\; \GG\zv_o \psx\; d^3\!x\quad -\quad{\dis 1\dover 4\pi}
{\intl_{\zS}} \psx\zv_o\GG\; d^3\!x  \cr
\noalign{\vskip8pt}
\quad -\quad {\hbox{$1$}\over\hbox{$4\pi$}} {\intl_{\zS}}
\;\GG{\dis x\over\dis (x^2+a^2)^\half} \; \zv_1 \psx   \; d^3\!x
\; +\;{\hbox{$1$}\over\hbox{$4\pi$}} {\intl_{\zS}}
{\displaystyle x\psx\over\hbox{$(x^2+a^2)^\half$}} \;\zv_1\GG\; d^3\!x
\end{array}
\eb

\< We can rewrite the fourth integrand as a total derivative and integrate
by parts.

\beg
\begin{array}{l}
{\dis 1\dover{4\pi}} \intl {\dis x \psx \dover (x^2+a^2)^\half}
\; \zv_1 G(x^0(x^1),x^1,x^2,x^3) \; d^3\!x  \cr
\noalign{\vskip5pt}
=
{\dis 1\dover{4\pi}} \intl {\dis x \psx \dover (x^2+a^2)^\half}
\; \biggl\{ {\dis d\dover dx}+[\zv_x(x^2+a^2)^\half]\zv_o \biggr\}
G(-\sqrt{x^2+a^2},x) \; d^3\!x  \cr
\noalign{\vskip5pt}
=
{\dis 1\dover{4\pi}} \intl {\dis x\psx\dover (x^2+a^2)^\half}
\; \biggl\{{\dis d\dover dx}+{\dis x\dover(x^2+a^2)^\half}\zv_o \biggr\}
G(-\sqrt{x^2+a^2},x) \; d^3\!x  \cr
\noalign{\vskip5pt}
=-
{\dis 1\dover{4\pi}} \intl\GG\;{\dis d\dover dx}
\biggl\{ {\dis x \psx\dover (x^2+a^2)^\half}\biggr\} \; d^3\!x
+{\dis 1\dover{4\pi}}\intl{\dis x^2\psx\dover(x^2+a^2)}\zv_o\GG d^3\!x\cr
\noalign{\vskip5pt}
=
 {\dis 1\dover{4\pi}}\intl {\dis x^2\psx\dover(x^2+a^2)}\; \zv_o\GG \; d^3\!x
-{\dis a^2\dover{4\pi}}\intl{\dis\GG\psx\dover(x^2+a^2)^\ahalf3}\; d^3\!x \cr
\noalign{\vskip5pt}
+{\dis 1\dover{4\pi}}\intl\GG{\dis x^2\dover(x^2+a^2)}\zv_o\psx\;d^3\!x
-{\dis 1\dover{4\pi}}\intl\GG{\dis x\dover(x^2+a^2)^\half}\zv_1\psx\;d^3\!x
\end{array}
\eb

\<We can rewrite the solution then as

\beg
\begin{array}{l}
\zc(\z9x_o)\quad =\quad
{\dis 1\dover{4\pi}}\intl\GG
\left\{ 1+{\dis x^2\dover(x^2+a^2)}\right\} \zv_o\psx\;d^3\!x   \cr
\noalign{\vskip5pt}
+{\dis 1\dover{4\pi}}\intl
\psx \left\{ {\dis x^2\dover(x^2+a^2)}-1\right\}\;\zv_o\GG \; d^3\!x  \cr
\noalign{\vskip5pt}
-{\dis 1\dover{2\pi}}\intl\GG{\dis x\dover(x^2+a^2)^\half}\;\zv_1\psx\;d^3\!x
\cr
\noalign{\vskip5pt}
-{\dis a^2\dover{4\pi}} \intl{\dis\GG\psx\dover(x^2+a^2)^\ahalf3}\; d^3\!x  \cr
\end{array}
\eb

\<Simplifying,

\beg
\begin{array}{l}
\zc(\z9x_o)\quad =\quad \cr
{\dis 1\dover{4\pi}}\intl\GG
\biggl\{ 2-{\dis a^2\dover(x^2+a^2)}\biggr\}\zv_o\psx\;d^3\!x
-{\dis a^2\dover{4\pi}} \intl{\dis\GG\psx\dover(x^2+a^2)^\ahalf3}\; d^3\!x  \cr
\noalign{\vskip5pt}
 -{\dis a^2\dover{4\pi}}\intl
 {\dis \psx\dover(x^2+a^2)}\;\zv_o\GG \; d^3\!x
-{\dis 1\dover{2\pi}}\intl\GG{\dis x\dover(x^2+a^2)^\half}
\;\zv_1\psx\;d^3\!x \cr
\end{array}\label{int0}
\eb

\<The third integral contains a derivative on $\GG$ that we need to
remove.

\<Now consider the general formula,

\beg
\intl_D\zd(f(x))\;g(x) dx=\sum_i{\dis g(x_i)\over\dis|\zv f/\zv x|}_{\dis x_i}
\eb
\smallskip

\<where the $x_i$ are the roots of the function $f$ in the range of integration
$D$.  We may then express a 2 dimension integral of the same form,

\beg
\begin{array}{cc}
 \intl_D \zd(f(x,y))\;g(x,y) dy dx
 &= \intl_{D\cap D(f)} \!\sum_i{\dis g(x,y_i(x))\over\dis
 \phantom{x}\big\vert \zv f/\zv y\big\vert}_{\!{\dis y_i}(x)}
\!dx\cr
 \end{array}\label{form1}
  \eb

\<where the functions $y_i(x)$ are the roots of $f(x,y_i)=0$.
$D(f)$ is the domain of the function $f$ where $\zv f/\zv y$ is defined, and
$D(x)$ is the original domain of the integration in $x$.
As a variation of the above identity, we also need to calculate the integral
of the derivative of the delta function with respect to its argument (in this
case, the function $f$).

\beg
\begin{array}{l}
 \intl_D \zd'(f(x,y))\;g(x,y) dy dx =
-\intl_D \!{d\over dy}
 \left\{\dis g(x,y)\over\dis\phantom{x}\zv f/\zv y \right\}
 \zd(f(x,y)) \;dy\;dx\cr
 \noalign{\vskip7pt}

 = -\!\intl_{D(x)\cap D(f)}\!\sum_i
\left[{\zv\over\zv y} \left\{\dis g(x,y)\dover \phantom{x}\zv f/\zv y\right\}
      {\dis 1\dover |\zv f/\zv y|}\right]_{\!{\dis y_i}(x)}
\;dx\cr
 \noalign{\vskip7pt}

 = -\!\!\intl_{D(x)\cap D(f)}\!\!\sum_i
\left[ \left\{
 {\dis g'(x,y)\dover \phantom{x} f'}-{\dis g(x,y)f''\dover \phantom{x} (f')^2}
\right\}
      {\dis 1\dover |f'|}\right]_{y_i(x)}\!\!
\;dx\cr
\end{array}\label{form2}
\eb

\<Where the prime refers to $y$ partial derivatives.

\<We can use \eqn{form1} and \eqn{form2} to reduce \eqn{int0}.
We begin by changing to polar coordinates,

\beg
\begin{array}{l}
\left.\begin{array}{l}
X = x-x_o = R \cos\zy \cr
Y = y-y_o = R \sin\zy\cos\zf \cr
Z = z-z_o = R \sin\zy\sin\zf \cr
\end{array} \right\} \z9R = (T,R)     \cr
T = t-t_o                             \cr
\end{array}
\label{polar}
\eb

\<Then the $\GG$ has the form

\beg
 \GG = {\displaystyle \zd(t_o-t-\dx0) \over\dis \dx0}
     = {\displaystyle \zd(-T-R)\over\dis R}
     = {\displaystyle \zd(T+R) \over\dis R}  = \GR,
 \eb

\<and its derivative is

\beg
\zv_o \GG = {\dis 1\dover R}\; \zv_T\zd(-T-R)
          = {\dis 1\dover R}\; \zv_R\zd(T+R)
\eb

Rewriting the third integral in \eqn{int0}, we have

\beg
\begin{array}{l}
-{\dis a^2\dover{4\pi}}\intl_\zS
\psx{\dis 1\dover(x^2+a^2)}\;\zv_o\GG\; d^3\!x \cr
\noalign{\vskip5pt}
=- {\dis a^2\dover{4\pi}}\intl_\zS\psr {\dis 1\dover\rcosa}{\dis 1\dover R}
\;\zv_T\zd(T+R) \; R^2 \sin\zy\ d\!R\ d\zy\ d\zf \cr
\noalign{\vskip5pt}
=- {\dis a^2\dover{4\pi}}\intl_\zS\psr{\dis 1\dover\rcosa}
\;\zd'(T+R)\; R\ d\!R\, \sin\zy\ d\zy\ d\zf \cr
\noalign{\vskip5pt}
= {\dis a^2\dover{4\pi}}\intl_\zS
\left[ {\zv \over \zv R}
\Bigl\{{\dis R\zvar \dover\rcosa}{\dis 1 \dover \zv f/\zv R} \Bigr\}
{\dis 1 \dover |\zv f/\zv R|}\right]_{f=0}
\!  \sin\zy\ d\zy\ d\zf, \cr
\end{array}\label{int1}
\eb

\<where we have applied \eqn{form2} and

\beg
\begin{array}{ll}
f(R,T) = R + T
   =& R - t_o - \rcosa^\half = 0 \cr
\imp& R - t_o = \rcosa^\half     \cr
\imp& R^2\sinq - 2 \cosn R - (a^2+x_o^2-t_o^2) = 0 \cr
\end{array}
\label{quadra}
\eb

\<We must take the positive root to solve for $R$,

\beg
\begin{array}{ll}
R(\zy)
&= {\dis \cosn\dover\sinq} +
{\dis\sqrt{\cosn^2 + \sinq (a^2+x_o^2-t_o^2)} \dover\sinq} \cr
\noalign{\vskip7pt}
&= {\dis \cosn\dover\sinq} +
{\dis\sqrt{\cost^2 + a^2\sinq} \dover\sinq} \cr
\noalign{\vskip7pt}
 &\equiv
 {\dis\cosn\dover\sinq} +
   {\dis A(\zy)\dover\sinq}\cr
\end{array}
\label{radius}
\eb

\<where $A(\zy)\equiv \sqrt{\cost^2 + a^2\sinq}$.
The derivatives of $f$ are

\beg
{\dis\zv f\dover\zv R} = 1-{\dis \rcos\cosy\dover\rcosa^\half}
\; ;\;
{\dis \zv^2f\dover \zv R^2} ={\dis -a^2\cosq\dover\rcosa^\ahalf3}
\eb

\<Evaluate these at $f=0$ ($R = R(\zy) $).

\beg
\begin{array}{l}
{\dis \zv f\dover\zv R}\Bigr\vert_0 = {\dis R-\rcos\cosy-t_o\dover R-t_o }
\equiv{\dis\A\dover R-t_o}\cr
\noalign{\vskip5pt}
{\dis \zv^2f\dover\zv R^2}\Bigr\vert_0 ={\dis -a^2\cosq\dover (R-t_o)^3} \cr
     \end{array}\label{der1}
      \eb

\<To complete the evaluation of \eqn{int1}, note first that

\beg
\begin{array}{l}
 {\dis d\dover dR} \Bigl\{{\dis R\zvar \dover\rcosa}\Bigr\} =
 {\dis x_o^2+a^2- R^2\cosq \dover\rcosa^2}\;\zvar   \cr
\noalign{\vskip8pt}
-{ \dis R\rcos\cosy\dover\rcosa^\ahalf3}\;\zvar_o
\quad + R\biggl\{
{ \dis \zvar_1\cosy +\zvar_2\cossin +\zvar_3\sinsin \dover\rcosa}
\biggr\}
 \end{array}
 \eb

\<Taking the $f=0$ limit and using \quad $\rcos\cosy=R-t_o-A$
(it is clear from equation \eqn{quadra} that $R-t_o\geq a$):

\beg
\begin{array}{l}
+{\dis a^2\dover{4\pi}}\intl_\zS\psx{\dis 1\dover(x^2+a^2)}\;\zv_o\GG\; d^3\!x
= -{\dis a^2\dover{4\pi}}\intl_\zS
\biggl\{ {\dis x_o^2\!+\!a^2\!-\!R^2\cosq\dover\rt^2\A^2} \biggr\}\zvar\dom\cr
\noalign{\vskip8pt}
\quad
+{\dis a^2\dover{4\pi}}\intl_\zS
\biggl\{
 { \dis R(R-t_o-\A)\dover\rt\A^2}
\biggr\}\zvar_o \dom
-{\dis a^2\dover{4\pi}}\intl_\zS
\biggl\{ {\dis a^2 R\cosq\dover\rt^2\A^3} \biggr\}
\zvar\dom\cr
\noalign{\vskip8pt}
\quad
-{\dis a^2\dover{4\pi}}\intl_\zS
R \biggl\{
{ \dis \zvar_1\cosy +\zvar_2\cossin +\zvar_3\sinsin \dover\A^2}
\biggr\}\dom .\cr
\end{array}\label{int2}
\eb

\<$\dom$ is the differential solid angle $\sin\zy\ d\zy\ d\zf$.
The solution \eqn{int0} becomes

\beg
\begin{array}{l}
\zc(\z9x_o)\quad = \cr
\noalign{\vskip7pt}\quad
{\dis 1\dover{4\pi}}\intl_\zS
\biggl\{ {\dis2\rt\dover\A}-{\dis a^2\dover\rt\A} \biggr\}
R\zvar_o\;\dom   \cr
\noalign{\vskip7pt}\quad
-{\dis a^2\dover{4\pi}}\intl_\zS
\biggl\{
{\dis R\dover\rt^2\A}
\biggr\}\zvar\dom
-{\dis a^2\dover{4\pi}}\intl_\zS
\biggl\{
{\dis R^2\cosq\!-\!x_o^2\!-\!a^2\dover\rt^2\A^2}
\biggr\}\zvar\dom\cr
\noalign{\vskip8pt}\quad
+{\dis a^2\dover{4\pi}}\intl_\zS
\biggl\{
 { \dis R(R-t_o-\A)\dover\rt\A^2}
\biggr\}\zvar_o \dom \cr
\noalign{\vskip8pt}\quad
+{\dis a^2\dover{4\pi}}\intl_\zS
R \biggl\{
{ \dis \zvar_1\cosy +\zvar_2\cossin +\zvar_3\sinsin \dover\A^2}
\biggr\}\dom\cr
\noalign{\vskip8pt}\quad
+{\dis a^2\dover{4\pi}}\intl_\zS
\biggl\{ {\dis a^2 R\cosq\dover\rt^2\A^3} \biggr\}
\zvar\dom
 -{\dis 1\dover{2\pi}}\intl_\zS {\dis R\rcos\dover\A}\;\zvar_1 \dom     \cr
\end{array}\label{int3}
\eb

\<Regrouping this in terms of \zvar and its derivatives,

\beg
\begin{array}{l}
\zc(\z9x_o)\quad =\quad \cr
\noalign{\vskip7pt}
{\dis 1\dover{4\pi}}\intl_\zS
\biggl\{
-{\dis a^2 (R^2\cosq\!-\!x_o^2\!-\!a^2)\dover\rt^2\A^2}
-{\dis a^2R\dover\rt^2\A}
+{\dis a^4R\cosq\dover\rt^2\A^3}
\biggr\}
\zvar \dom \cr
\noalign{\vskip7pt}
+{\dis 1\dover{4\pi}}\intl_\zS
\biggl\{
 {\dis 2R\rt\dover\A}
-{\dis a^2R\dover\rt\A}
-{\dis a^2R(R-t_o-\A)\dover\rt\A^2}
\biggr\}
\zvar_o \dom    \cr
\noalign{\vskip7pt}
-{\dis 1\dover{4\pi}}\intl_\zS
\biggl\{
 {\dis 2R\rcos\dover\A}
-{\dis 2a^2R\cosy\dover\A^2}
\biggr\}
\zvar_1 \dom \cr
\noalign{\vskip7pt}
+{\dis 1\dover{4\pi}}\intl_\zS
\biggl\{
 {\dis a^2R\cossin\dover\A^2}
\biggr\}
\zvar_2 \dom
+{\dis 1\dover{4\pi}}\intl_\zS
\biggl\{
 {\dis a^2R\sinsin\dover\A^2}
\biggr\}
\zvar_3 \dom    \cr
\noalign{\vskip7pt}
\end{array}
\eb

\<which we rewrite as

\beg
\zc(\z9x_o)\quad =\quad
\dom\biggl\{ I_1 +I_2+ I_3 +I_4+ I_5 +I_6+ I_7 +I_8+ I_9 +I_{10} \biggr\}
\label{int4}
\eb

We have just reduced the solution to integrals over a finite angular
domain. In the following section we state and prove that the integrands are
everywhere finite, leading to a finite solution.

\subsection{Existence Theorem for the Scalar Field}

In this section we state and prove the following:

\begin{theorem} Let there be
$C^2$ data on the hypersurface
\beg\zS =\{(x,y,z)|(y,z)\in (-\infty,\infty) \;;\; t=-\sqrt{x^2+a^2} \}\eb
and zero data at $\scri^-\cup J^+(\zS)$,
denoted by $\zvar$ and $\zv_t\zvar\equiv\zvar_o$ that satisfy the following:

\beg
\zvar\; = O(R^{-q})
\;;\;\zv_\zm\zvar\;= O(R^{-p}),
\;;\; h_m :\zvar,\zvar_o=\zvar,\zvar_o
\label{data2}
\eb

\<where $q=\ahalf3$ and $p=2$,
and $h_m$ is the translation operator given in equation \eqn{trans}. Then
pointwise convergent solution for the wave equation $\zv_\zm\zv^\zm\zc=0$ on
Minkowski spacetime is,

\beg
\zc(\z9x_o) =
\opi\intl_{\zS} \GG\ddel_\za \psx dV^\za,
\eb

\<and its pullback by $\zp_n^{-1}$, where $\pi_n$ is the restriction of
$\pi$ to $D_n$, gives a convergent solution on the Gott Spacetime.
\end{theorem}

 One expects the solution to converge since as one moves to arbitrarily large
values of $y$, the identification maps $A_\pm$ of equation \eqn{id1}
push the data to exponentially
larger values of $x^\pm$, in which case the original restrictions on the
data for large $x$ provide even greater convergence. If one does not move
arbitrarily far in the $y$ direction, the original convergence of the data
suffices to guarantee a finite integrand.
The following proof demonstrates this explicitly.

\<{\bf Proof of Theorem:} Equation \eqn{int4} expresses the
solution in terms of angular integrals.
For each to be finite, it is sufficient
that its integrand be finite for all values of $\zy$ and $\zf$.

\begin{casex}
: $\z9x_o$ lies outside the lightcone of the origin.
\end{casex}

\vskip-1.2truein
\vbox{
\begin{figure}
\epsfig{file=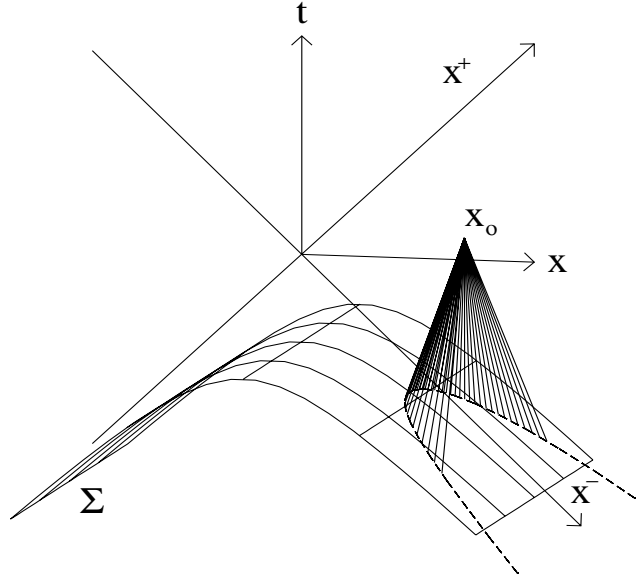,height=5.4in,width=6.0in}
\vskip-.6truein
\caption{The domain of integration for the solution consists of the
intersection of the past lightcone centered at $x_o$ with the initial data
surface $\zS$. The intersection becomes infinitely distant as one
approaches the $x$ axis.}
\label{init1}
\end{figure}
}

Although  $R(\zy)$ is divergent at $\zy=0$ (since the intersection of the
light cone with the hyperbolic sheet extends to infinity on the $x$ axis),
$R(\zy)$ is finite at $\zy=\pi$.
Without loss of generality, we take the region $-x_o<y_o<x_o\; ;\; x_o>0$.
\bigskip

\beg
\begin{array}{ll}
R(\zy) &= {\dis (x_o\cos\zy+t_o)\dover\sin^2\!\zy} +
{\dis\sqrt{(x_o\cos\zy+t_o)^2 + \sin^2\!\zy (a^2+x_o^2-t_o^2)}
\dover\sin^2\!\zy} \cr
\noalign{\vskip10pt}
       &= {\dis (x_o\cos\zy+t_o)\dover\sin^2\!\zy} +
          {\dis |x_o\cos\zy+t_o|\dover\sin^2\!\zy}
\dis\sqrt{1+\sin^2\!\zy{\dis (a^2+x_o^2-t_o^2)\dover (x_o\cos\zy+t_o)^2}}\cr
\noalign{\vskip10pt}
       &\simeq {\dis (x_o\cos\zy+t_o)\dover\sin^2\!\zy} +
          {\dis |x_o\cos\zy+t_o|\dover\sin^2\!\zy}
\left(
1+\dis\half\sin^2\!\zy{\dis(a^2+x_o^2-t_o^2)\dover(x_o\cos\zy+t_o)^2} +\cdots
\right),
\end{array}
\eb

\<hence,

\beg\simeq\dis\half\, |t_o-x_o|\, {\dis (a^2+x_o^2-t_o^2)\dover(t_o-x_o)^2},\eb

\<which is clearly finite for the choice $-x_o<y_o<x_o\; ;\; x_o>0$.
\<This can be seen geometrically in figure \ref{init1}.
A symmetric analysis holds for the case \hbox{$x_o<y_o<-x_o\; ;\; x_o<0$}.

Analysis of the asymptotic form of the function $R(\zy)$ for $\zy\sim 0$
shows that it is $O(\zy^{-2})$. The asymptotic form of $A(\zy)$ for small
$\zy$ is

\beg
\begin{array}{ll}
A(\zy\sim 0)
       &= (x_o\cos\zy+t_o)
\dis\sqrt{1+\sin^2\!\zy{\dis (a^2+x_o^2-t_o^2)\dover (x_o\cos\zy+t_o)^2}}\cr
\noalign{\vskip5pt}
       &={\dis (x_o\cos\zy+t_o)}
\left( 1+O(\zy^2) \right) \simeq  \;(x_o+t_o).\cr
\end{array}
\eb

\<It is now clear that $\;\zv f/\zv R\;$ in \eqn{der1} is non-zero in
the domain of integration, and we only need to worry about the behavior of
$R(\zy)$.
\<Note that only the derivative terms of \zvar are potentially divergent.
The term $A$ is always non-zero and finite and $\rt$ is positive definite but
divergent at $\zy = 0$.

\<For $\zf\simeq 0$, $m$ will be positive, and for $\zf\simeq\pi$, $m$ will be
negative. The polar coordinate transformation centered at $\z9x_o$
is given by

\beg
\begin{array}{l}
x = x_o + R \cos\zy \cr
y = y_o + R \sin\zy\cos\zf \cr
z = z_o + R \sin\zy\sin\zf \cr
\end{array}
\label{polar1}
\eb

First consider large radii with $y>0$,
corresponding to $\zy\simeq 0$ and $\cos\zf>0$. We proceed by finding limits
on the asymptotic form of $R$.

\beg
\begin{array}{ll}
R(\zy\sim 0)
       &= {\dis 1\dover\sin^2\!\zy} (x_o\cos\zy+t_o)
\Biggl\{1+
\sqrt{1+\sin^2\!\zy{\dis (a^2+x_o^2-t_o^2)\dover(x_o\cos\zy+t_o)^2}}\Biggr\}\cr
\noalign{\vskip10pt}
 \geq &
{\dis 2\dover\sin^2\!\zy} (x_o\cos\zy+t_o) > {\dis K'\dover\zy^2}
\end{array}
\label{asympr}
\eb

\<where we have set $K'\equiv(x_o+t_o)$.
These relations give a restriction on $\zy$:

\beg
x_o\cos\zy+t_o > K'/2 \imp \zy <
\cos^{-1} \Bigl( {\dis x_o-t_o\dover 2x_o} \Bigr) \eb

\<A similar calculation gives that

\beg R(\zy) < {\dis 6K'\dover\zy^2} \label{rlim},\eb

\<with $\zy$ sufficiently small so that the radical term in \eqn{asympr}
is less than $\sqrt{2}$. Under these circumstances,

\beg {\dis K'\dover\zy^2}< R(\zy) < {\dis 6K'\dover\zy^2} \eb.

Next, let's find the limits of $m$
in terms of $\zy$.
{}From \eqn{polar1} we can write the condition for being
in the $m^{th}$ two-cell:

\beg
\begin{array}{l}
(4m-2)Y_0\leq y \leq (4m+2)Y_0\cr
\noalign{\vskip10pt}
\imp  (4m+2)Y_0 > y_o + R \sin\zy\cos\zf > R\sin\zy\cos\zf\cr
\noalign{\vskip10pt}
\imp  m > {\dis R\sin\zy\cos\zf\dover 4Y_0} -\dis\half
	>\dis\eighth{\dis K_o\cos\zf\dover Y_0} {\dis 1\dover\zy}     \cr
\end{array}
\label{largem}
\eb

\<where we again used \eqn{asympr}. For the previous condition to hold,

\beg
\half < {\dis\sin\zy\dover\zy}-{\dis Y_0\zy\dover K'\cos\zf}
\qu\imp\qu \zy \leq \half\left[
{\dis 1\dover 3!}+{\dis Y_0\dover K'\cos\zf}
\right]^{-1}
\eb

We need to find the asymptotic form of the radius restricted to the
unit two-cell in order to find the form for the initial data.
Define $x_\zs$ as the value of $x$ when translated back to $D_o$ by $h_n$.
Writing the Lorentz transformation for coordinates on the
hyperboloid $t=-\sqrt{x^2+a^2}$, the asymptotic form for $x_\zs$ becomes

\beg
\begin{array}{ll}
|x_\zs|& = |x \cosh(4m\za) -\sqrt{x^2+a^2}\qu \sinh(4m\za)|\cr
\noalign{\vskip4pt}
&= x\cosh(4m\za)[\sqrt{1+{a^2\over x^2}}\qu\tanh(4m\za)-1]\cr
\noalign{\vskip4pt}
&\geq x\cosh(4m\za)\txt[(1+\fourth{a^2\over x^2})\Bigl(1-3{\ext{-8m\za}\over
1 +\ext{-8m\za}}\Bigr)-1]\cr
\noalign{\vskip4pt}
&\geq x\cosh(4m\za)[\txt\fourth{a^2\over x^2}-\eighth{a^2\over x^2}  ]
 \geq \dis\eighth {a^2\over x}\cosh(4m\za) \cr
\noalign{\vskip4pt}
& \geq \dis{a^2\over 16R} \ex{\txt\half{K_o\cos\zf\over Y_0}
{\za\over\zy}}
\geq\dis{a^2\zy^2\over 96K'}\ex{\txt\half{K_o\cos\zf\over Y_0}{\za\over\zy}}
\cr
\end{array}
\eb

\<where \eqn{largem} and \eqn{asympr} has been used. Then

\beg
R_\zs =  \sqrt{x_\zs^2 + y_\zs^2 + z_\zs^2}
\geq\dis{a^2\zy^2\over 96K'}\ex{\txt\half{K_o\cos\zf\over Y_0}{\za\over\zy}}.
\eb

\<For initial data that satisfies

\beg
\zvar\leq\dov{K}{R^{q}}\qu,\qu\zv_\zm\zvar\leq\dov{K_\zm}{R^{p}},
\eb

\<as in \eqn{data1}, we have

\beg
\zvar_\zs\leq
\dis{96^{q}KK'^{q}\over a^{2q}\zy^{2q}}
\ex{\txt-\ahalf{q}{K_o\cos\zf\over Y_0}{\za\over\zy}}
\qu ,\qu
\zv_t\zvar_\zs\leq
\dis{96^{p}K_oK'^{p}\over a^{2p}\zy^{2p}}
\ex{\txt-\ahalf{p}{K_o\cos\zf\over Y_0}{\za\over\zy}} .
\label{limit1}
\eb

\<Similarly, for $\cos\zf<0$, $m$ will be negative implying

\beg
\begin{array}{ll}
x_\zs& = x \cosh(4m\za)[1 +\sqrt{x^2+a^2}\;\tanh(4|m|\za)]\cr
\noalign{\vskip4pt}
&\geq x\cosh(4m\za) \geq x\ex{4|m|\za} \cr
\end{array}
\eb

\<The initial data on $\zs$ therefore obeys

\beg
\zvar_\zs\leq
\dis{K\zy^{2q} \over K'^{q}}
\ex{-\txt\ahalf{q}{K_o|\cos\zf|\over Y_0}{\za\over\zy}}
\qu ;\qu
\zv_t\zvar_\zs\leq
\dis{K_o\zy^{2p} \over K'^{p}}
\ex{-\txt\ahalf{p}{K_o|\cos\zf|\over Y_0}{\za\over\zy}}
\label{limit2}
\eb

\<The values $q=\ahalf3$ and $p=2$ were found in section \ref{initial}.
The conclusion is that for $\cos\zf\neq 0$, the data prescribed on the unit
two-cell is exponentially damped and gives convergent angular
integrals in \eqn{int4}, since all the other terms are polynomial in $\zy$.

The case where $\cos\zf=0$ is handled separately. In this case, we will
be inside the unit two-cell and we rely on the usual asymptotics of the
solution as in \eqn{asympr}:

\beg R_\zs \geq {\dis K'\dover\zy^2} \eb

\<We need only look at the solution \eqn{int4} and consider the worst case
divergences near $\zy = 0$. This leaves the integrals,

\beg
\zc\simeq
{\dis 1\dover{4\pi}}\intl {\dis 2R^2\dover\A} \zvar_o \dom
-{\dis 1\dover{4\pi}}\intl {\dis 2R^2\dover\A} \zvar_1
\sin\zy\cos\zy\dif\zy\dif\zf
\label{asym1}
\eb

\<Analyzing the first integrand, we ignore $A$ and constants which are finite,
so that the asymptotic form for the integrand,

\beg R^2\zvar_o\sin\zy\leq R^2{\dis K_o\dover R^{2}}\zy=K_o\zy\to 0 \eb

\<is convergent in the angular integration, and  we have used \eqn{asympr}
and \eqn{data1}. The second integral in \eqn{asym1} similarly gives

\beg R^2\zvar_1\sin\zy\leq R^2{\dis K_1\dover R^{2}}\zy=K_1\zy \to 0.\eb

\<Here we have used \eqn{data1}.
The conclusion is that in the region defined by $x_o>0$ and $-x_o<y_o<x_o$,
all the integrals in \eqn{int4} are finite.

\begin{casex}
: $\z9x_o$ lies on the past lightcone of the origin.
\end{casex}

In this the case, we will show that the intersection of $\zS$ with the
past light cone of $\z9x_o$
extends a finite distance in $y$ which leads to the same asymptotic
behavior as in the Minkowski case.

On the past lightcone, we set $t_o=-x_o$. It is assumed that $x_o>0$ so
that $\zy$ near zero is the only singular region needed. Of course, the
case $x_o<0$ is completely analogous.
\<From \eqn{quadra}, the asymptotic form for the radius is

\beg
\begin{array}{ll}
      & R^2\sinq + 2 x_o (1\!-\!\cos\zy) R - a^2 = 0 \cr
\noalign{\vskip10pt}
\imp  & R(\zy) = {\dis 1\dover\sinq}
\Bigl\{\txt-x_o(1\!-\!\cos\zy)+\sqrt{x_o^2(1\!-\!\cos\zy)^2+a^2\sinq}\Bigr\}\cr
\noalign{\vskip10pt}
      &        = {\dis 1\dover\sin\zy}
\Bigl( -x_o{\txt\tan\tht} \sqrt{a^2+(x_o \tan\tht)^2} \Bigr)
\equiv {\dis 1\dover\sin\zy}\;K(\zy) \cr
\end{array}
\label{quadra1}
\eb

\<Since $K(\zy)$ monotonically increases as $\zy$ goes to zero,
$\exists \ze>0$ such that $\forall \zy <\ze$, $\txt R(\zy)\geq
{\txt K_\ze\tover\sin\zy}
\geq {\txt K_\ze\tover\zy}$, where $K_\ze\equiv K(\ze)$.
{}From \eqn{radius}, the asymptotic form for the $A(\zy)$ can be found as

\beg
\begin{array}{c}
A(\zy) = [a^2\sinq +x_o^2(1\!-\!\cos\zy)^2]^\half
= a\sin\zy
\Bigl[1+\bigl({{\txt x_o}\tover a}\bigr)^2
\Bigl({\txt1\!-\!\cos\zy\tover\sinq}\Bigr)^2
\Bigr]^\half \cr
\geq a\sin\zy \geq {\txt a\zy\tover 2}
\end{array}
\label{quadra2}
\eb

\<It is assumed that we are in a regime where $\sin\zy\geq\zy/2$.
It is readily verified that

\beg
{\txt a\zy\tover 2}\leq a\sin\zy \leq A(\zy)\leq
a\bigl[1+\txt\bigl({\txt x_o\tover a}\bigr)^2\bigr]^\half\sin\zy\leq aK_a\zy
\eb

\<Where $K_a \equiv \bigl[1+\bigl({\txt x_o\tover a}\bigr)^2 \bigr]^\half $.
{}From the two prior calculations, we have bounds on $R$,

\beg {\txt K_\ze\tover\zy} \leq R(\zy)\leq {\txt aK_a\tover\sin\zy}\leq
{\txt 2aK_a\tover\zy}\eb

\<The distance in the $y$ direction is finite as follows,

\beg
\begin{array}{l}
t=-\sqrt{x^2+a^2} \qu;\qu t_o-t = R \cr
\imp\qu t_o + \sqrt{x^2+a^2}
- \Bigl[(x-x_o)^2+(y-y_o)^2+(z-z_o)^2\Bigr]^\half\cr
\imp\qu
(y-y_o)^2 = t_o^2 + x^2 + a^2 + 2t_o\sqrt{x^2+a^2}-(x-x_o)^2-(z-z_o)^2.\cr
\end{array}
\eb

\<Setting \, $t_o=-x_o$\,,

\beg
\begin{array}{l}
(y-y_o)^2 = a^2 + 2x_ox - 2x_o\sqrt{x^2+a^2}-(z-z_o)^2 \cr
\imp\qu (y-y_o)^2 \leq a^2 - (z-z_o)^2 \cr
\end{array}
\eb

\<We see that the bounding values for $y$ are given by \hbox{$y = y_o \pm a$}.
This implies that the initial data on the unit cell is transformed as in
equation \eqn{lorentz} for a fixed value $m'>0$.
We want to show that there exists a value of $\zy_c$ such that for
$\zy<\zy_c$,

\beg
\begin{array}{l}
\zvar_\zs\leq KR_\zs^{\txt-\ahalf3}
\leq K\biggl({\dis \zy\dover K_{m'}}\biggr)^{\ahalf3} \cr
\zv_\zm\zvar_\zs\leq K_\zm R_\zs^{-2}
\leq K_\zm\biggl({\dis \zy\dover K_{m'}}\biggr)^{2}
.\cr
\end{array}
\label{coneasym}\eb

\<for constant $K_{m'}$. First take $m'>0$. We find that if

\beg \zy<{\txt K_\ze\tover a\ex{4m'\za}} \eb

\<then

\beg
\begin{array}{ll}
x_\zs& =  x \cosh(4m'\za) -\sqrt{x^2+a^2}\qu \sinh(4m'\za)\cr
\noalign{\vskip4pt}
& =   x[\cosh(4m'\za) -\sqrt{1 +{\scr a^2\over\scr x^2}}\qu \sinh(4m'\za)]\cr
\noalign{\vskip4pt}
&=  \dis\ahalf{x}
\Bigl[\ex{4m'\za}+\ex{-4m'\za}-\Bigl(1+\half{\dis a^2\dover x^2}-\cdots\Bigr)
\Bigl(\ex{4m'\za}-\ex{-4m'\za}\Bigr) \Bigr]\cr
\noalign{\vskip4pt}
&\geq
\dis\ahalf{x}\ex{-4m'\za} \Bigl[-\half{\dis a^2\dover x^2}\ex{8m'\za}+2 \Bigr]
\geq\dis\ahalf{x}\ex{-4m'\za}.
\end{array}
\eb

\<The last condition implies that for $\zy<{\txt K_\ze\tover a\ex{4m'\za}}$

\beg
R_\zs> {\dis K_\ze\dover4\zy} \ex{-4m'\za}\equiv {K_{m'}\dover\zy}.
\eb

\<A similar calculation for $m'<0$ gives

\beg
\zy<{\txt K_\ze\tover a}\ex{-4m'\za} \qu;\qu
R_\zs>
{\dis K_\ze\dover4\zy}
\dis \ex{-4m'\za}\equiv {K_{m'}\dover\zy}.
\eb

In both cases, the lower bound for $R_\zs$ will have the same
$\zy^{-1}$ behavior, which will suffice to show convergence of equation
\eqn{int4}.
\<The asymptotic form of $x_\zs$ restricted to the unit cell
behaves like a constant times the unrestricted $x$ and the asymptotic
analysis for either case will be identical.
The initial data on the lightcone thus obeys \eqn{coneasym}.

We now proceed to show that the integrals are finite for each term.
The ten terms in the integrand of \eqn{int4} have the asymptotic form

\beg
I_1\leq
{\dis a^2  R^2\cosq\dover R^2\A^2} \zvar\sin\zy
\leq {\dis 4Ka^2\dover(a\zy)^2}\zcs\zy
\leq {\dis 4K\dover K_{m'}^{\ahalf3} }\zy^{\half} \to 0
\eb

\beg
I_2
\leq{\dis a^2\dover  R\A} \zvar\zy
\leq{\dis 2Ka^2\dover a\zy}
{\dis\zy\dover K_\ze}
\zcs\zy
 \leq{\dis 2aK\dover K_\ze K_{m'}^\ahalf3}\;\zy^{\ahalf5} \to 0
 \eb

\beg
I_3
\leq{\dis a^4\dover  R\A^3} \zvar\zy
\leq{\dis8Ka\dover \zy^3} {\dis\zy\dover K_\ze}\zcs\zy
\leq{\dis8Ka\dover K_\ze K_{m'}^\ahalf3}\;\zy^{\half} \to 0
\eb

\beg
I_4
\leq{\dis 8R^2 \dover a\zy} \zcst\zy
\leq{\dis 8K_o\dover a\zy}
\biggl({\dis 2aK_a\dover\zy}\biggr)^2
\zcst\zy
\leq{\dis 16aK_aK_o\dover K_{m'}^2}
\eb

\beg
I_5\leq
{\dis a^2R\dover\rt\A}\sin\zy \zvar_o \leq
{\dis 2a^2K_o\dover a\zy}\zcst\zy\leq
{\dis 2aK_o\dover K_{m'}^2}\zy^{2}\to 0
\eb

\beg
I_6\leq
{\dis a^2R\dover\A^2}\zy \zvar_o \leq
{\dis (2aK_aK_o)\dover\zy^2}\zcst\leq
{\dis (2aK_aK_o)\dover K_{m'}^2}
\eb

\beg
I_7\leq
{\dis 4R^2\cos\zy\dover a}\zvar_1\leq
{\dis 4K_1\dover a}
\biggl({\dis\zy\dover K_{m'}}\biggr)^{2}
\biggl({\dis 2aK_a\dover\zy}\biggr)^2 \leq
{\dis 16aK_a^2K_1\cos\zy\dover K_{m'}^2}
\eb

\beg
I_8\leq
{\dis 2a^2R\cosy\dover\A^2}\zvar_1\sin\zy\leq
{\dis 16aK_aK_1\dover\zy^2}
\biggl({\dis\zy\dover K_{m'}}\biggr)^{2}\leq
{\dis 16aK_aK_1\dover K_{m'}^2}
\eb

\beg
I_9\leq
{\dis a^2R\cos\zf\sinq\dover\A^2}\zvar_2 \leq
{\dis 8aK_aK_2\dover K_{m'}^2}
\zy\to 0
\eb

\<where we have used the relations \eqn{partial}, \eqn{coneasym},
and \eqn{asympr}. The last integral in \eqn{asym1} has the same behavior as
$I_9$ and was omitted.
\eject

\begin{figure}
\epsfig{file=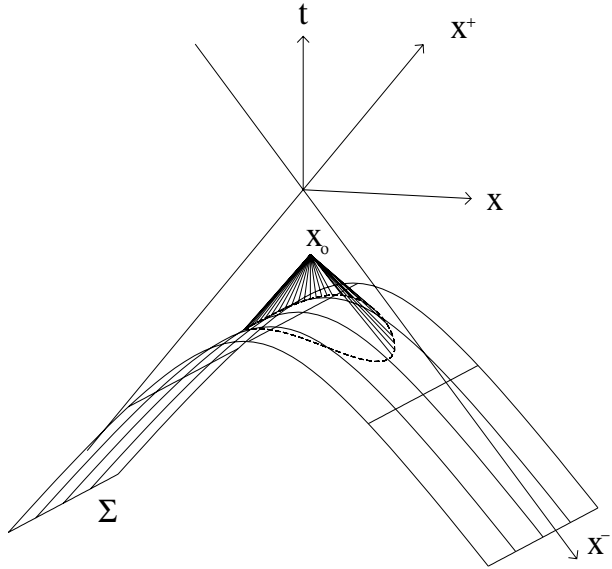,height=4.0in,width=5.0in}
\caption{The domain of integration for the solution when
$x_o$ lies inside the past lightcone of the origin.
The intersection covers a finite region of $\zS$. }
\label{init2}
\end{figure}

\vbox{
\begin{casex}
: $\z9x_o$ lies inside the past lightcone of the origin.
\end{casex}

In this case, the intersection of $\zS$ with the past light cone of $\z9x_o$
lies in a finite region (see figure \ref{init2}) where the data is $C^\infty$
and finite, so that the integral \eqn{int4} is finite and $C^\infty$.
}

\begin{casex}
{:} {$\z9x_o$ lies inside the future lightcone of the origin, $J^+(0)$.}
\end{casex}

By cases 1,2, and 3, there exists a constant time surface $\zS_a$, external to
$J^+(0)$, with data $\zvar_a$. This implies that a
solution exists inside $J^+(0)$.\fim

\appendix

\section{Isometry Group Elements of H}
\label{isogroup}
For some calculations it is required to know the isometry elements that
carry a point $X\in\mbar$ to the $n^{th}$ cell away from its
origination. The following presents a detailed calculation of this.
It will also be demonstrated that the group elements are significantly
simplified for translation of an even number of unit cells.

 Let $h_{mn}\in H$ be the group element that carries $X_n$ from the $m^{th}$
level
to the $n^{th}$ level in \mbar. We can construct $h_{mn}$ from the projection
map $\pi$ by projecting first onto \M\ and then applying $\pi^{-1}$ restricted
to the $m^{th}$ level;

\beg
\begin{array}{ll}
h_{mn}&= \pi^{-1}_m\comp\pi_n
       = L^{m} P^m  K^{-\hbox{$s_m$}} K^{\hbox{$s_n$}} P^n L^{-n}    \cr
&= L^{m} P^m  K^{\hbox{$s_n$}\!-\hbox{$s_m$}}P^n L^{-n} \cr
&= L^{m}  K^{\hbox{$s_{mn}$}\!-m} P^{n+m} L^{-n} \cr
&= L^{m-\hbox{$s_{mn}$}}  K^{\hbox{$s_{mn}$}\!-m} P^{n+m}\cr
\end{array}
\eb

\<The above relation applied to a point $X= (X^+,X^-,Y)\in D_n$ gives

\beg
\begin{array}{l}
 h_{mn} X
 =  L^{m-s_{mn}}  K^{s_{mn}\!-m} P^{n+m} X \cr
 =  L^{m-\hbox{$s_{mn}$}}
\left(\begin{array}{ccc}
 e^{2(s_{mn}\!-m)\za} &                      0 \       &\ 0  \cr
    0                 &\!\! e^{\!-2(s_{mn}\!-m)\za} &  0  \cr
    0                 &                      0         &  1  \cr
    \end{array} \right)
   P^{n+m} X
\end{array}\label{iso1}\eb

\<where we have used the definitions \eqn{pi} and \eqn{pi_inv}.
This can be written for several particular cases. First take $m+n$ even. Then
$s_{mn} = n$ and $P^{m+n} = 1$, which gives

\beg
 \begin{array}{l}
 h_{mn} X
  =  L^{m-n}  K^{n-m}  X =  L^{m-n}
\left(\begin{array}{ccc}
 e^{2(n-m)\za} &       0        &\ 0\ \cr
    0          & e^{-2(n-m)\za} &  0  \cr
    0          &        0       &  1  \cr
    \end{array} \right) X \cr
 =  (e^{2(n-m)\za}X^+,e^{-2(n-m)\za}X^-,Y+2(m-n)Y_0)
  \end{array}
  \label{even}
    \eb

\<Next take $m+n$ odd ($s_{mn} = -n$, $P^{m+n} = P$)

\beg
 h_{mn} X
  =  L^{m+n}  K^{-n-m} P X
  =  -\xvec(e^{-2(n+m)\za}X^-,e^{2(n+m)\za}X^+,{ Y\!-\!2(m\!+\!n)Y_0})
  \label{odd}
    \eb

\<Now assume that $X$ is at the $n^{th}$ level (which it must be
for the previous formulae to apply), and that $m=n+l$. Then even $n+m=2n+l$
implies even $l$,
and we can rewrite
equations \eqn{even} and \eqn{odd} with $h_{mn}=h_l$,

\beg
 h_{l} X
  =  L^l  K^{-l} X
  =  L^l
\left(\begin{array}{ccc}
 e^{-2l\za} &    \  0 \     &\ 0\ \cr
 \  0\         & e^{2l\za}  &  0  \cr
    0          &    0       &  1  \cr
    \end{array} \right) X
 =  \xvec(e^{-2l\za}X^+,e^{2l\za}X^-,Y\!+\!2lY_0) .
  \label{even1}
    \eb

\<For odd $l$, odd $n+m$ implies odd $l$, so we cannot express $h_{mn}$
as $h_{m-n}$, which still depends on the initial level that we start from,

\beg
 h_{l} X
 =  L^{2n+l}  K^{-2n-l} P \xvec(X^+,X^-,Y)
 =  -\xvec(e^{-(4n+2l)\za}X^-,e^{(4n+2l)\za}X^+,{Y\!\!-\!(4n\!+\!2l)Y_0})
  \label{odd1}
    \eb
\section{$Y-Z$ Independent Solutions }

One may wish to determine if solutions to the wave equation on the Gott
spacetime

\beg\squ\phi = (\del_x^2 +\del_y^2 + \del_z^2 - \del_t^2 ) =0\phi\eb

\<can be both $y$ and $z$ independent. It will be shown in this section
that this is impossible. If it were, the wave equation becomes

\beg \squ \phi(x^+,x^-) = \del_+ \del_- \phi(x^+,x^-) = 0 \label{2dequ} \eb

\<This equations implies a solution of the form

\beg \phi(x^+,x^-) = f(x^+) + g(x^-) \eb

\<Now apply the boundary condition \eqn{rec2} at $y=Y_0$,

\beg \begin{array}{lrc}
&\phi(x^+,x^-)     \quad =& \phi(-e^{-2\za}x^-,-e^{2\za}x^+) \cr
\imp\quad & f(x^+)+g(x^-)\quad =& \quad f(-e^{-2\za}x^-)+g(-e^{2\za}x^+) \cr
\imp\quad & f(x^+)-g(-e^{2\za}x^+)\quad =&\quad f(-e^{-2\za}x^-)-g(x^-) \cr
\imp\quad & f(x^+)    \quad  =& k_1 +  g(-e^{2\za}x^+) \cr
   \end{array}\label{fn1} \eb

\<Similarly for the boundary condition \eqn{boost1} at $y=-Y_0$,

\beg \begin{array}{lrc}
&\phi(x^+,x^-)     \quad =& \phi(-e^{2\za}x^-,-e^{-2\za}x^+) \cr
\imp\quad & f(x^+)    \quad  =& k_2 +  g(-e^{-2\za}x^+) \cr
\end{array}\label{fn2} \eb

\<Equations \eqn{fn1} and \eqn{fn2} combine to give a recursion relation

\beg \begin{array}{lc}
 f(-e^{-4\za}x^+) =& g(-e^{-2\za}x^+) + k_1 = f(x^+)+ (k_1-k_2) \cr
\imp &f(x^+) = f(e^{-4n\za}x^+)+ n (k_1-k_2) \cr
 \end{array}\label{fn3} \eb

\<where $n$ is integer. For $f$ to be finite, we require that $(k_1-k_2)=0$.
Taking the derivative of $f$

\beg \begin{array}{lc}
{\hbox{$\dif{f}(x^+)$}\over\hbox{$\dif x^+$}}=& e^{-4n\za}f'(e^{-4n\za}x^+)\cr
  \imp f'(e^{-4n\za}x^+) &= e^{+4n\za} f'(x^+) \cr
   \end{array}\label{fn4} \eb

\<If we fix some value for $x^+$, then $f(x^+)$ is a constant say $a$.
Then we have

\beg f'(e^{-4n\za}x^+) = a e^{+4n\za} \to \infty \eb

\<which shows that if $f$ takes on a non zero value anywhere, it becomes
infinite near $x^+=0$.

  This shows that the solution may not be simultaneously independent of $y$
and $z$ unless it is identically zero on the whole space.
\section{The Quotient Space and the Group Action}
\label{gappen}

The following theorem applies for a general smooth differential operator $k$,
manifold M, and base space B. The operator $h$ is properly discontinuous
as verified in the next section of the appendix.

\begin{theorem}
Let \M\ be a manifold, \H\ a properly discontinuous group acting on M.
Let $\B=\M/\H$\ be a quotient spaced with projection
$\pi:\M\mapsto\B$. Define $L(\M)$ as the space of smooth functions
on \M. \H\ acts on $L(\M)$ via pull-backs: For $h\in \H$, $x\in\M$, and
$\zc\in L(\M)$,

\beg (h^*\zc)(x)=\zc(h(x)).\eb

\<Let $\z4k$ be a sufficiently smooth linear operator on L(\M) that is
invariant under \H:
\beg h^*\comp \z4k\comp (h^{-1})^*= \z4k\qu\qu \forall\; h\in\H.\eb
Then, the push-forward of $\z4k$ defines a smooth linear operator $k$ on
$L(\B)$.
\end{theorem}

\<Proof:Take $x\in\B, \z4x_1, \z4x_2\in \pi^{-1}(x), \z4x_1\neq\z4x_2$.
Then we can find open neighborhoods $U$, $V_1$, and $V_2$ such that
for the following homeomorphisms,

\beg\begin{array}{cc}
f_1: &V_1\qu \lrtu{\pi_{|V_1}}\qu U   \cr
f_2: &V_1\qu \lrtu{\pi_{|V_2}}\qu U   \cr
 g : &V_1\qu \lrtu{h_{|V_1}}  \qu V_2,
\end{array}\eb

\<the following diagram commutes:
\medskip

\begin{figure}
\vbox{
\hfil
\epsfig{file=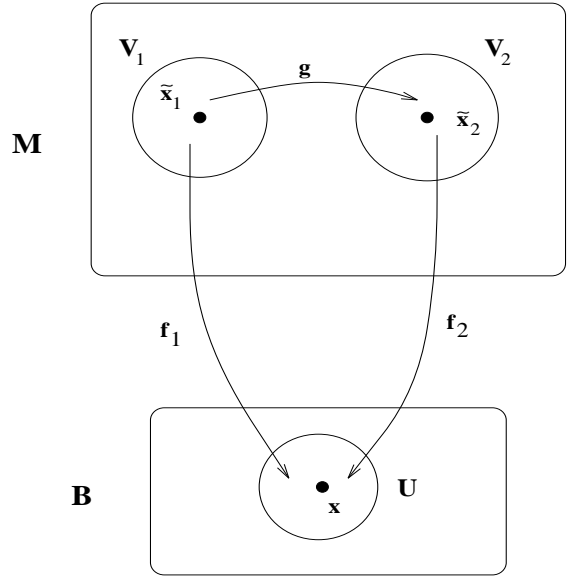,height=3.0in,width=3.0in}
\hfill
}
\bigskip
\caption{Commutative diagram for a quotient space B. The maps $f_1$, $f_2$,
and $g$ are shown, along with the neighborhoods $U$, $V_1$, and $V_2$ about
the points $x$, $x_1$, and $x_2$.}
\end{figure}

\<In other words, $f_1=f_2\comp g$.

Define in $L(U)$ the operators

\beg\begin{array}{cc}
k_1= & f_1^* \comp\z4k\comp (f_1^{-1})^* \cr
k_2= & f_2^* \comp\z4k\comp (f_2^{-1})^* \cr
\end{array}\eb

\<We have that

\beg\begin{array}{cc}
     &f_1=f_2\comp g          \cr
\imp &f_1^*= f_2^*\comp g^*   \cr
     &f_1^{-1}= g^{-1}\comp f_2^{-1} \cr
\imp &(f_1^{-1})^*= (g^{-1})^* \comp (f_2^{-1})^* \cr
\end{array}\eb

\<This implies that

\beg\begin{array}{rc}
k_1\quad= & f_1^* \comp\z4k\comp (f_1^{-1})^* \cr
= &\qu f_2^*\comp\underbrace{g^*\comp\z4k\comp(g^{-1})^*}\comp(f_2^{-1})^* \cr
= &\qu f_2^*\qu\comp\qu \z4k\qu \comp\qu(f_2^{-1})^* \cr
= &\! k_2,
\end{array}\eb

\<where we have used the assumption that $\z4k$ is invariant under H.
The definition of $k_i$ does not depend on the choice of preimage and is
therefore unique. This defines the operator $k$ on $L(\B)$.\fim

\section{Proof that $h$ is Properly Discontinuous}

We wish to show that $h$ is properly discontinuous.
Since $\pi$ is a projection, every point of $x\in \M$ has an open NBHD
$U_x \in \M$ such that $\pi^{-1}$ maps $U_x$ to a disjoint union of
neighborhoods in \mbar. This is useful to show the following:

\bigskip
\<1. Show that each $\xbar \in \mbar$ has a open nbhd \vbar such that
\beg h_n(\vbar) \cap \vbar = \nill \qu \forall\qu n\neq 0.\eb

 Pick a nbhd \vbar of $\xbar\in\mbar$ such that $\pi(\xbar)=x$.
Let $U_x$ be a nbhd of $x\in\M$ for which $\pi^{-1}(U_x)$ is a disjoint
union of open nbhds in \mbar. Let
$\vbar' \equiv \pi^{-1}(\vbar)\cap\vbar$ . Then

\beg \sum_{n\neq 0} h_n(\vbar')\cap\vbar' = \nill \eb

since each image $h_n(\vbar')$ is a pre-image of $\pi^{-1}$ that is
disjoint.

\bigskip
\<2. For $\xbar_1,\xbar_2\in\mbar$ with
$\xbar_1\neq h_n(\xbar_2)\qu\forall\qu n$, there are nbhds $\vbar_1$
and $\vbar_2$ with \hbox{$h_n(\vbar_1)\cap\vbar_2=\nill \qu\forall\qu n$.}

 First take both $\xbar_1$ and $\xbar_2$ to lie interior of all lines
$ y=(2n+1)Y_0 $. Then there is a minimum $y$ distance between
$h_n(\xbar_1)$ and $\xbar_2$ for some $n'$. Let $\epsilon$ be defined as

\beg
\epsilon\equiv MIN(d(h_{n'}(\xbar_1),\xbar_2)/4,d_1).
\eb

\<where $d(.,.)$ is the positive distance between points and $d_1$
is the closest distance to any line $ y=(2n+1)Y_0 $ from
either of $\xbar_1$ or $\xbar_2$.
If we take open balls $B_\ze(\xbar_1)$ and
$B_\ze(\xbar_2)$, centered at $\xbar_1$ and $\xbar_2$ respectively
and radius $\ze$, then the diameter of $h_n(B_\ze(\xbar_1))$ remains the same
as $B_\ze(\xbar_1)$ since $h_n$ is an isometry. Since
$h_n(\vbar_1)\cap\vbar_2$ can only be nonempty if the $y$ coordinate
are in the same cell ($n=n'$), we conclude that the intersection is always
empty.

 Next take $\xbar_1$ on a line $ y=(2m+1)Y_0 $ and $\xbar_2$ interior to
all such lines. Again, there is a $n'$ such that $h_{n'}(\xbar_2)$ is
is closest in $y$ to $\xbar_1$. Again we take

\beg \epsilon\equiv MIN(d(h_{n'}(\xbar_1),\xbar_2)/4,d_1).  \eb

\<except that $d_1$ is now the closest distance to any of the lines
$ y=(2n+1)Y_0 $ from $\xbar_2$. In this case as well, let
$\vbar_1=B_\ze(\xbar_1)$ and $\vbar_2=B_\ze(\xbar_2)$. Then,

\beg h_n(\vbar_1)\cap\vbar_2=\nill \qu\forall\qu n \eb

If both $\xbar_1$ or $\xbar_2$ lie on some lines $ y_1=(2n_1+1)Y_0 $ and
$ y_2=(2n_2+1)Y_0 $, then there is some $h_n$ for which
$h_n(y_1)=y_2$. Since it is assumed that $\xbar_1$ and $\xbar_2$ are not
related by any $h_n$, we are guaranteed that $h_{n}(\xbar_1)\neq \xbar_2 $,
so that

\beg \epsilon\equiv d(h_{n}(\xbar_1),\xbar_2)/4 \eb

\<will give a $\vbar_1=B_\ze(\xbar_1)$ and $\vbar_2=B_\ze(\xbar_2)$ with

\beg h_n(\vbar_1)\cap\vbar_2=\nill \qu\forall\qu n \eb
\newpage
\centerline{\bf Acknowledgements}

 I am grateful John L. Friedman for his many suggestions and insights.
I would also like to thank Jorma Louko, Steve Winters,
Nick Papastamatiou, Eli Lubkin, Nick Stergioulas, and T. C. Zhao.
Computer time provided by National Center for Supercomputing Applications.
This work was partially supported by NFS grant No. PHY-91-05935 and a U.S.
Department of Education Fellowship. Sustenance provided by Lakefront
Brewery.

\end{document}